\documentclass[manuscript]{acmart}
\usepackage{xcolor}
\pagecolor{white}
\usepackage{mathtools,xparse}
\usepackage{multirow}
\usepackage{graphicx,subfig}
\usepackage[export]{adjustbox}
\usepackage{caption}
\usepackage{amsthm} 
\usepackage{thmtools}

\usepackage{hyperref}
\usepackage{cleveref}
\usepackage[compact]{titlesec}
\titlespacing{\subsection}{0pt}{10pt}{1pt}
\titlespacing{\subsubsection}{0pt}{10pt}{5pt}
\titlespacing{\section}{0pt}{10pt}{1pt}

\usepackage{thm-restate}
 
\declaretheoremstyle[%
  spaceabove=6 pt,%
  spacebelow=6 pt,%
  bodyfont=\itshape,
  headfont=\normalfont \scshape ,%
  postheadspace=0.2em,%
  qed=\qedsymbol,%
  headpunct = .
]{mystyle}

\declaretheoremstyle[%
  spaceabove=6 pt,%
  spacebelow=6 pt,%
  bodyfont=\normalfont,
  headfont=\normalfont \scshape ,%
  postheadspace=0.2em,%
  qed=\qedsymbol,%
  headpunct = .
]{proofstyle}
\declaretheorem[name={Lemma},style=mystyle,numbered,]{Lemma}

\declaretheorem[name={Proof},style=proofstyle,unnumbered,]{Proof}
\declaretheorem[name={Definition},style=mystyle,numbered,]{Definition}

\usepackage{proof-at-the-end}

\usepackage[belowskip=0pt,aboveskip=0pt]{caption}

\AtBeginDocument{%
  }

\setcopyright{rightsretained}
\acmJournal{TORS}
\acmYear{2024} \acmVolume{1} \acmNumber{1} \acmArticle{1} \acmMonth{1}\acmDOI{10.1145/3674157}

\acmJournal{TORS}
\acmPrice{}
\acmISBN{}

\usepackage{enumitem}
\setlist{nosep,after=\vspace{\baselineskip}}
\begin{document}

\title{Evaluating Impact of User-Cluster Targeted Attacks in Matrix Factorisation Recommenders}


\author{Sulthana Shams}
\affiliation{%
  \institution{Trinity College Dublin}
  \country{Ireland}}
\email{sshams@tcd.ie}

\author{Douglas Leith}
\affiliation{%
  \institution{Trinity College Dublin}
  \country{Ireland}}
\email{Doug.Leith@tcd.ie}

\begin{abstract}
In practice, users of a Recommender System (RS) fall into a few clusters based on their preferences. In this work, we conduct a systematic study on user-cluster targeted data poisoning attacks on Matrix Factorisation
(MF) based RS, where an adversary injects fake users with falsely crafted user-item feedback to promote an item to a specific user cluster. We analyse how user and item feature matrices change after data poisoning attacks and identify the factors that influence the effectiveness of the attack on these feature matrices. 
We demonstrate that the adversary can easily target specific user clusters with minimal effort and that some items are more susceptible to attacks than others. Our theoretical analysis has been validated by the experimental results obtained from two real-world datasets. Our observations from the study could serve as a motivating point to design a more robust RS.
 
\end{abstract}




\maketitle

\section{Introduction}
Recommender Systems (RS) have become increasingly important in recent years due to the explosion of online data and the need for personalized recommendations. They help users discover new products, services, or content based on their interests and past behaviors, leading to increased engagement and revenue for businesses.

Users in a RS can often be grouped by their interests, and indeed this observation motivates many advertising strategies \cite{FloC, clustering_ads_1, clustering_ads_2}. Recent research proposes exploiting such clustering of users in RS  for their 'hiding in the crowd' privacy benefits \cite{BLC, FloC}. However, it is well known that RS are susceptible to data poisoning attacks \cite{fun_and_profit, attack_survey_1}. Social media platforms have received attention for discriminatory and predatory targeting of user communities. Introducing false product reviews and creating fake social media accounts to promote misinformation are examples of how adversaries use RS to influence users. RS recommends items by learning user’s preferences and contributes to shaping and developing new opinions and interests, thus influencing user behavior. Therefore, selective exposure to misinformation is a significant concern in RS.

In poisoning attacks, an adversary adds fake profiles with carefully assigned ratings for selected items and attempts to target an item to promote/demote it by increasing/decreasing the item’s rating in the system. This study explores the impact of data poisoning attacks targeting specific user clusters in Matrix Factorisation (MF) based RS.  

Given a RS with $n_1$ users and $n_2$ items, MF is a popular recommendation technique that works by decomposing the original sparse user-item interaction matrix $R \in \mathbb{R}^{n_1\times n_2}$ into two low-rank matrices $U \in \mathbb{R}^{d\times n_1}$ and $V \in \mathbb{R}^{d\times n_2}$  where their row dimension $d$ is typically much less than $n_1$ and $n_2$ \cite{MF}. This allows us to uncover latent features and the association between users and items to these latent features in the $d$ dimensional latent space.  
The two low-rank matrices $U$ and $V$ associate a user’s inclination towards the latent features and an item’s degree of membership towards those latent features respectively. Their matrix product $U^TV$ approximates $R$ and predicts the missing entries of the original sparse matrix. We call $U$ the user-feature matrix and $V$ the item-feature matrix. 
When fake ratings are introduced into a RS, the feature matrices $U, V$ undergo changes. It is crucial to analyse the nature and extent of these changes to conduct further research on the robustness of MF-based RS. Our study aims to investigate the impact of poisoning attacks on the feature matrices $U$ and $V$, an aspect that has not been previously investigated in the literature. In particular, we investigate attacks targeting specific user groups and how the changes to $U$ and $V$ matrices enable these attacks. 

The standard MF approach is to update both $U$ and $V$ matrices when new ratings enter RS. To investigate how $U$ and $V$ are individually affected by the attacks and how their behaviors contribute to the propagation of targeted attacks in the RS, we consider updates to each matrix separately in response to the newly entered fake ratings \cite{MF_ours}, i.e. i) Hold $V$ constant and update $U$ to study the changes to $U$ after attacks,  ii) Hold $U$ constant and update $V$ to study the changes to $V$ after attacks.  Our findings indicate that the impact of the attack is pronounced when $V$ is updated in response to the introduction of fake ratings.

The effectiveness of attacks on feature matrices can be influenced by the choice of target items. Most work in the literature \cite{data_poison_study_1, influence_based, graph_based, deep_learning_based, adv_attacks_1, targeted_attack} studies the effect of attacks on randomly chosen target items or on items with a small number of ratings. In this paper, we take a broader view and consider various choices of target item to identify the factors contributing to attack effectiveness on the $U, V$ matrices by analyzing how these matrices change after an attack on these target items. We conclude that certain items are more susceptible to attacks than others. Specifically, we have found that items with a lower number of ratings in the target cluster are particularly vulnerable. This is because the feature vectors for these items in matrix $V$ can be easily manipulated, making them more susceptible to attacks. So the distribution of user and item latent features plays a significant role in the effectiveness of attacks targeting a user group.


Many recent works approach data poisoning attacks from an optimization perspective, requiring a high level of knowledge about the RS \cite{influence_based, adv_attacks_1, NMF_attack}.  For our work, we revisit more straightforward attack strategies \cite{fun_and_profit, attack_survey_1, attack_survey_3} involving users rating the target item with a large rating to promote it and the rest of the items (filler items) with scores that mimic the rating distribution of the targeted user cluster. We show that this low-knowledge attack is enough to promote an item in a target cluster. 

\vspace{1mm}
Our main contributions are as follows:
\begin{itemize}

 \item  We provide a systematic study of data poisoning attacks targeted at a group of users and identify the factors contributing to the effectiveness of attacks in MF-based RS.
  \item  
  We analyse the individual effects of attacks on latent feature matrices $U$ and $V$ and explore how they contribute to the propagation of targeted attacks in a MF-based RS. Studies investigating the role of latent feature matrices are new to the literature.
     
  \item We illustrate our findings with real-world datasets and show that a simple attack strategy using limited knowledge of user preferences suffices to target a specific user group precisely.
\end{itemize}
    
\section{Related Work}

The benefits of clustering users in RS are discussed in \cite{BLC, fast_cold_start, clustering_users_1, clustering_users_2}. For example, \cite{BLC} proposes enhancing privacy by assigning the users to clusters and providing cluster-wise recommendations rather than for each individual user 
and \cite{fast_cold_start, clustering_users_1} exploits user clusters for improving cold-start recommendation.    Recently Google introduced a new algorithm \cite{FloC} which uses such clustering of users. Known as Federated Learning of Cohorts, or FLoC, it assigns individual browsers to larger groups of users based on their browsing history to enable privacy-oriented personal recommendations. FloC allows a form of ``hiding in the crowd'' privacy. 

The impact of data poisoning attacks where fake users are injected in RS with carefully crafted user-item interaction has been studied extensively. For surveys on attack models and the robustness of RS algorithms, see~\cite{fun_and_profit, attack_survey_1, attack_survey_2, attack_survey_3}. Studies such as \cite{deep_learning_based, factorisation_based,graph_based,influence_based} focus on modelling attacks specific to the type of RS, e.g. \cite{influence_based} studies attacks in MF-based RS by proposing to model fake users similar to true users who are influential to the recommendations, \cite{deep_learning_based,graph_based} propose data poisoning attacks for deep learning based RS and graph-based RS respectively.  

Very few studies explore the impact of data poisoning attacks targeting specific user clusters in MF-based RS. The closest work to the present paper is probably~\cite{targeted_attack, adv_attacks_1}.  In \cite{adv_attacks_1}, Generative Adversarial Networks (GAN) generate fake users whose distribution is close to that of the true users. They assume that the adversary has knowledge of the recommender's model and algorithm and aims at adversarial training of RS to build robust recommendation models (for other adversarial robustness studies, see~\cite{adv_attacks_1,adv_attacks_2,adv_attacks_survey}).  In~\cite{targeted_attack}, they propose to promote an item to a group of users by injecting fake ratings and social connections in a factorisation-based social RS. They assume that fake users can quickly establish social connections with true users and model the attack as a bi-level optimisation problem. 

In the above studies, attacks are modeled as an optimisation problem. They require a good deal of information about the RS to be carried out, i.e. a rather powerful adversary.   However, observations from  
these studies motivate us to revisit the more straightforward average-attack strategy and study their effect on MF-based RS.  For example, in \cite{graph_based}, the attacker generates fake users using a graph-based RS, and the target RS uses MF.  The differences between their attack and the existing standard attack strategies are not very large despite the high level of knowledge assumed by the adversary. Similar observations are made in \cite{NMF_attack}, where the authors study data poisoning attacks on deep learning RS models. They generate fake profiles by randomly sampling a sub-matrix from real users who have rated sufficient items to generate fake users close in distribution to the real users. They compare their attack strategy to standard attack models such as a random attack, average attack etc, and find that the effectiveness of the attack is only marginally increased compared to the lower knowledge and simpler average-attack strategy. 

Recently there have been a range of improvements and expansions to the fundamental MF. While the inner product approach, which linearly combines latent features, has its merits, it may not be sufficient to capture the complex structure of user interaction data. For instance, \cite{NCF} explores the use of deep neural networks for learning the interaction function from data. In NCF (Neural Collaborative Filtering) user and item feature vectors are fed into a multi-layer neural architecture to map these latent vectors to prediction scores. Similarly \cite{deep_MF} proposes a novel MF model with neural network architecture as a simple nonlinear generalization of MF.

However, recent findings outlined in \cite{MFvsNCF} suggest that utilizing a dot product could serve as a more effective default choice for combining embeddings compared to learned similarities using Neural Collaborative Filtering (NCF). The proposal comes with several positive implications: (1) The research becomes more relevant for the industry because MF models are applicable for real time recommendation where as MLP similarity is not applicable for real time top-N recommenders. The dot product enables fast retrieval through established nearest neighbor search algorithms (2) Dot product similarity simplifies modeling and learning (no pre-training, no need for large datasets). This simplicity not only facilitates experimentation but also enhances overall understanding. 

These insights, focusing on the foundational technique of MF, prompt a deeper exploration of the MF strategy under attack.

\section{User-Cluster Targeted Poisoning Attack}

By grouping people into clusters with similar preferences, we can exploit the collective behaviours and preferences of users within each cluster. This approach allows for improved recommendations as users within the same cluster tend to exhibit higher degrees of similarity compared to the broader user population. This concept of user clustering in RS is not a new one. In \cite{CF_clustering, smoothing_clustering, plsa_clustering}, a range of techniques (K-means and Gibbs sampling, a cluster-based smoothing system, and a cluster-based latent semantic model respectively) are proposed which cluster users and items in an unsupervised manner to improve the accuracy of prediction and scalability problems. Netflix, for example, has adopted this strategy by segmenting its vast user base of over 93 million individuals into approximately 2,000 clusters based on taste \cite{netflix-geography-age-gender, taste_communities_netflix}.
More recently, Google's Privacy Sandbox initiative aims to address privacy concerns by categorizing users into interest groups based on their browsing behavior \cite{FloC, Fledge}. This approach provides a strong "hiding in the crowd" type of privacy for advertisers, publishers, and ad tech providers to deliver personalized ads.

In such a user cluster-based RS, clusters formed around shared traits or preferences become appealing targets for attackers. These attackers can take advantage of the cluster by entering and controlling it to alter recommendations and target content to a sizable user base. The study by \cite{segment_attack_analysis} was among the earliest to investigate the effects of attacks focused on a specific user segment of  RS. They demonstrate that such attacks are not spread evenly across all users; they are directed towards a specific subset of the user base. This strategy of limiting the impact eliminates suspicion and reduces the chance of being discovered.

Fake reviews flooding Amazon listings \cite{awsblog, facebookbusiness, amazon_fake_reviews, nytimes-amazon-fake-reviews, washington-amazon-fake-reviews}, discriminatory targeting in advertising \cite{facebook_ads, ads_google} and organised content suppression efforts \cite{washington-antivaccine} are just a few real-world examples of these attacks.
As a result, the advantages of clustering for privacy and advertising also raise worries about its possible role in aiding targeted attacks.

\subsection{User Cluster Based Recommendation Model}

In an MF-based recommender, the user-item rating matrix $R$ is first factorized approximately as $U^TV$, where matrices $U$ and $V$ have relatively small inner dimension $d$.  Each column in matrix $U$ is a length $d$ weight vector that captures a user's preferences, and each column in matrix $V$ is a length $d$ weight vector that captures the characteristics of one item. 

The predicted rating matrix given by the inner product $U^TV$ gives the individual predicted rating per user per item. 

We depart from the usual setup by assuming that users belong to distinct groups. We have a set $\mathcal{G}$ of user groups, and each user belongs to one group $g \in \mathcal{G}$ such that users belonging to a group have similar preferences. In our study, we use $k$-means to cluster columns of matrix $U$ so that users with similar preferences (i.e. weight vectors) are clustered together. However, other clustering algorithms might also be used, e.g. the approach in \cite{BLC}.

Let $\tilde{U}_g \in \mathbb{R}^{d\times1}$ for $g=1,\cdots,|\mathcal{G}|$ be the preference vector associated with a group of users and gather these features to form matrix $\tilde{U} \in \mathbb{R}^{d \times |\mathcal{G}|}$. Each column in $\tilde{U}$ is a length $d$ weight vector that captures a group's preferences. For users belonging to a group $g$ with weight vector $\tilde{U_g}$ and items with weight vectors in $V$, the predicted rating matrix is the inner product $\tilde{U_g}^TV$, i.e., users belonging to the same group have the same predicted rating value per item.

\subsection{Attack Model}

We assume that the adversary knows the cluster-wise mean rating for every item in the system. For the type of attacks that we focus on, there is a \emph{target item} that the attacker is interested in promoting in a target cluster and a set of \emph{filler items} that are rated to ensure that the fake users preferences share some similarity to that of the true users in the target cluster. The target item is given the maximum rating to try to promote it in the system and the filler items ensure that fake users can correctly enter the targeted group. 

The assumptions on the adversary's knowledge are reasonable since such cluster-wise information may be inferred from online databases that publicly display the average user ratings of items and metadata such as user's age, gender, etc. (e.g., movie databases, online shopping databases, etc.). For example, popular platforms like Amazon and Goodreads publicly display user ratings and reviews for items or books they've engaged with. Attackers can leverage general domain knowledge about the target item to find products with similar genres or attributes that align with their objectives. By identifying users who have rated these related items highly within the RS, the attacker can identify potential targets and subsequently emulate the choice patterns of these target users by aggregating ratings from these selected users for other items.

\section{Study of Attack Effects on Feature Matrices $U$ and $V$}\label{update_approaches}

The standard approach to obtain $U, V$ given user-item ratings is to minimize the sum of squared error over the set of all observed ratings, i.e. 

\begin{equation}\label{costfunc}
 \min_{U,V}\sum_{(i,j)\in \mathcal{O}} (R_{i,j}-U_i^TV_j)^2  + \lambda\left(\sum_{i}\lVert U_i\rVert^2+\sum_{j}\lVert V_j\rVert^2 \right)
\end{equation}

Where $\mathcal{O}$ is the set of (user, item) rating pairs, $R_{i,j}$ is the rating of item $j$ by user $i$, $U_i$ and $V_j$ are the column vectors that describe the preferences associated to user $i$ and item $j$ respectively; $\lVert.\rVert$ is the Euclidean norm and $\lambda$ is a regularisation weight. We re-cluster these $U_i$ values to obtain matrix $\tilde{U}$ of group vectors.

When fake users arrive and start rating items, we study changes in $U, V$ by updating only one of the feature matrices in response to newly entered ratings and keeping the other constant. In particular, we have the following approaches to incorporate the new ratings:

\begin{enumerate}

\item \textit{Fix $V$, update $U$: }
RS holds $V$ constant and updates $U$ for all users after the attack. Keeping V constant, the cost function in equation \ref{costfunc}, becomes for every user $i$, a convex function of $V$. The solution to this least squares problem is

\begin{align} \label{Vconst}
U_i  
     &=\left(\sum_{j \in \mathcal{V}(i)}{V_jV_j^T+\lambda I}\right)^{-1}\sum_{j \in \mathcal{V}(i)}V_jR_{i,j}   
\end{align}

where $\mathcal{V}(i)=\{j:(i,j) \in \mathcal{O}\}$ is the set of items rated by user $i$. 
Then re-cluster the $U_i$ values to obtain an updated matrix $\tilde{U}$ of group vectors.

\item \textit{ Fix $U$, update $V$: }
RS holds $U$ constant and updates $V$ after the attack. 
\begin{align} \label{Uconst}
V_j&=\left(\sum_{i \in \mathcal{U}(j)}{U_iU_i^T+\lambda I}\right)^{-1}\sum_{i \in \mathcal{U}(j)}U_iR_{i,j} 
\end{align}
where $\mathcal{U}(j)=\{i:(i,j) \in \mathcal{O}\}$ is the set of users rating item $j$.
\end{enumerate}

\subsection{Fix $V$, Update $U$}\label{varyU}
  
It can be seen from equation \ref{Vconst} that user vector $U_i$ depends on the item vector $V_j$ and rating value $R_{i,j}$ of all items $j \in \mathcal{V}(i)$ rated by user $i$. For any existing true user $i$ in RS, the addition of fake users neither changes the set $\mathcal{V}(i)$ of items rated nor $R_{i,j}$ and $V_j$ of items in $\mathcal{V}(i)$. So the weight vector $U_i$ is unchanged after an update provided $V$ is kept constant.

Let $U_f$ be the feature vector of a fake user $f$ entering the RS. $U_f$ is generated based on the set of items $\mathcal{V}(f)$ rated by the fake user and the ratings given to these. After generating $U_i$ values for all users (now includes fake users as well), the cluster-feature weights $\tilde{U}$ are updated using k-means. We also warm start the $k$-means clustering with the initial centers before the attack instead of using a random cluster center initialization.   This allows us to focus on regrouping due to adding fake users and avoid spurious group membership changes due to randomness in the initial conditions.

\subsubsection{Impact of Attack on Target Cluster Weight Vector $\tilde{U}_t$}\label{cluster_pop}
The new cluster center is the arithmetic mean of all user weights in the cluster. When fake users successfully enter the target cluster $t$ by generating feature vectors $U_f$ that are close to $\tilde{U}_t$, we update the cluster center to reflect the presence of these new users. The updated $\tilde{U}_t$ is the mean of the feature vectors of the existing true users and the newly added fake users. Therefore the predicted rating of target item $j^*$ in the target cluster given by $\tilde{U}_t^TV_{j^*}$ changes. Therefore, the effect of the attack on item $j^*$, in this case, is determined by the changes to the cluster center $\tilde{U}_t$ resulting from the inclusion of fake user weights. Based on this understanding, we expect the relative number of true users ($n$) and fake users ($m$) in the target cluster to play an essential role in the effectiveness of the attack. If $n>>m$, then the impact of fake users on the arithmetic mean $\tilde{U}_t$ is small.

\subsubsection{Impact of Attack on Non-Target Items and Non-Target Clusters}\label{leakage_items}
 The inner product $\tilde{U_{t}}^TV$ gives the predicted rating for all items in the target cluster. Specifically, we expect that, in the target cluster, items similar to $j^*$ (i.e: item feature vectors $V_j$ and $V_{j^*}$ have high correlation) will show a more significant change in the predicted ratings. In contrast, other items will show a lower change in the predicted ratings. 

Recall that any true user's feature vector remains unchanged after injecting fake users in RS. So we expect non-target cluster center $\tilde{U}_g$ for $g \in \mathcal{G}\setminus \{t\}$ to be unchanged if all fake users correctly enter target cluster $t$. Thus, keeping $V$ constant will isolate the attack effects on the target cluster alone if all fake users enter the target cluster.

\subsubsection*{}

Surprisingly, our analysis shows us that the injection of fake ratings has no effect on the user vector $U_i$ of all true users $i$ in the RS (when $V$ is kept constant). Also, the number of true users in target cluster $t$ relative to the number of fake users determines the attack's effect on $\tilde{U}_t$. 
It is $V_{j^*}$ that changes in response to injected fake ratings. Further, the post-attack changes are influenced by the number of true ratings to item $j^*$ from the target cluster rather than the total number of true users in the target cluster. We will study this in detail in the next section.

\subsection{Fix $U$, Update $V$} \label{varyV}

It can be seen from equation \ref{Uconst} that a target item vector $V_{j*}$ depends on user vector $U_i$  and rating $R_{i,j^*}$ of all users $i$ in set $\mathcal{U}(j^*)$ who provide rating to the item $j^*$. $V_{j^*}$ remains unaffected by the number of users in a cluster unless they also provide a rating to item $j^*$.

Let a fake user $f$ enter RS targeting item $j^*$. This implies that the set $\mathcal{U}(j^*)$ of users now also contains the fake users. Following this, fake user feature vector $U_f$ and rating $R_{f,j^*}$ are also used to determine $V_{j^*}$. Therefore, target item vector $V_{j^*}$ will show significant changes after the attack when many fake users enter RS. 

\subsubsection{Impact of Attack on Target Item Vector $V_{j^*}$} \label{V_iter}
Let $\hat{V}_{j^*}$ be the updated $V_{j^*}$ after attack.
Given the $k^{th}$ item-feature of $V_{j^*}$ represented by $v^{j^*}_k$ and $k^{th}$ user-feature of $\tilde{U}_t$ represented by $u^{t}_k$, the change in target item predicted rating in the target cluster after the attack, for this setup, is $$\tilde{U_t}^T\hat{V}_{j^*}-\tilde{U_t}^TV_{j^*}=\sum_{k=1}^d u^{t}_{k}(\hat{v}^{j^*}_{k}-v^{j^*}_{k})$$

 The change in predicted rating will increase when $\hat{v}^{j^*}_{k}-v^{j^*}_{k}=\gamma u^t_{k}$ for a positive constant $\gamma$. This is because, the inner product will then become $\sum_{k=1}^d u^{t}_{k}(\hat{v}^{j^*}_{k}-v^{j^*}_{k})=\sum_{k=1}^d \gamma (u^{t}_{k})^2$. We can see that, as $\gamma$ increases, product terms in the inner product become larger and positive, increasing the change in the predicted rating of the target item. 
Specifically, we define the following setup to study the transformation of $V_{j^*}$.
We noted from equation \ref{Uconst} that updating $V_{j^*}$ after the attack requires the $U_i$ values corresponding to all users who have provided ratings, i.e., true users and newly added $m$ fake users. Since users in a group have similar preferences, we set the $U_i$ values of true users to their corresponding cluster weight vector for this study. Given vector $U_f \in \mathbb{R}^{d \times 1}$ associated with fake user $f$ capturing the fake user's preferences,  we gather these vectors together to form matrix $X \in \mathbb{R}^{d \times m}$.  To improve the predicted rating of the target item in the target cluster $t$, columns in the $X$ matrix must be closer to $\tilde{U}_t$ than to other cluster feature vectors. For maximum changes to $V_{j^*}$, which corresponds to the maximum change in the predicted rating of the target item, we set $U_f$ equal to $\tilde{U}_t$ in our analysis.

We re-write equation \ref{Uconst} such that after adding a block of $m$ fake users, we can formulate updates to $V_{j^*}$ recursively. 

\begin{restatable}[\it{Proof in Appendix}]{Theorem}{Viterative}\label{V_iterative}
 Given $\hat{V}_{j^*}$ is the updated $V_{j^*}$ after attack. Let a block of $m$ fake users with feature vectors in $X \in \mathbb{R}^{d \times m}$ and target item fake ratings vector $y \in \mathcal{R}^{m \times 1}$ enter RS. Given $A^{-1}=\left(\sum_{i \in \mathcal{U}(j^*)}U_iU_i^T+\lambda I\right)^{-1}$ for set of all true users $\mathcal{U}(j^*)$ rating item $j^*$ and $K=\left(I+X^TA^{-1}X\right)^{-1}\left(y-X^TV_{j^*}\right)$, we can write,

\begin{align} \label{A_inverseU}
\hat{V}_{j*}-V_{j*} &= m \times K \times \left(A^{-1}\tilde{U}_t\right) 
\end{align}
\end{restatable}

\subsubsection*{Analysis of Vector $V_{j^*}$ Update Mechanism}
Using equation \ref{A_inverseU}, for any $k^{th}$ item-feature of $V_{j^*}$ given by $v^{j^*}_k$ and $k^{th}$ user-feature of $\tilde{U}_t$ given by $u^{t}_k$, we can write the updated $k^{th}$ item-feature $\hat{v}^{j^*}_k$ of $\hat{V}_{j^*}$ as
\begin{align}\label{feature_update}
    \hat{v}^{j^*}_{k}-v^{j^*}_k &= m \times K \left(a_{kk}u^{t}_{k}+\sum_{i=1,i\ne k}^d a_{ki}u^{t}_{i}\right)
\end{align}

where  $a_{kk}$ represents the diagonal and $a_{ki} \text{ for } \{i \mid i\ne k \text{ and } i \in 1,2,\cdots,d\}$ represents non-diagonal elements in the $k^{th}$ row of matrix $A^{-1}$. We can clearly observe from equation \ref{feature_update} that $\hat{v}^{j^*}_{k}-v^{j^*}_{k}$ is not a simple scaling of $u^{t}_k$ due to the presence of the second summation term. Therefore the transformation to any feature $v_k$ depends on the matrix $A^{-1}$ and target cluster-feature vector $\tilde{U}_t$.

We know that $A^{-1}$ is a Positive Definite (PD) matrix. \footnote{See Appendix \ref{derivingV}} Thus all diagonal elements $a_{kk}$ are positive.\footnote{See Lemma \ref{PSD_diagonalpositive} in Appendix} From this we can say that the term $a_{kk}u^{t}_{k}$ always has the sign of $u^{t}_{k}$ or in other words scales $u^{t}_k$. 
Equation \ref{feature_update} reduces to a simple scaling of $u^{t}_k$ only when $A^{-1}$ is a positive diagonal matrix causing the non-diagonal terms in any row of $A^{-1}$ to be zero i.e. the summation term $\sum_{i=1,i\ne k}^d a_{ki}u^{t}_{i}$ in equation \ref{feature_update} becomes zero giving $\hat{v}^{j^*}_{k}-v^{j^*}_k=m \times K \times a_{kk} \times u^{t}_{k}$.  

Let's take a closer look at the possible behaviors of equation \ref{feature_update} for any $k^{th}$ feature in item vector $V_j$. We know that $m, K$ are positive constants. So the overall sign and magnitude of the update is determined by the behaviour of term $a_{kk}u^{t}_{k}+\sum_{i=1,i\ne k}^{d} a_{ki}u^{t}_{i}$. We noted how $a_{kk}u^{t}_{k}$ is a term always having the sign of $u^{t}_k$, effectively acting as a scaling of $u^{t}_k$. 
Depending on the sign and magnitude of the term $\sum_{i=1,i\ne k}^d a_{ki}u^{t}_{i}$ relative to $a_{kk}u^{t}_k$, we can have the following behaviours:

\begin{itemize}
    \item case 1: if $sgn(a_{kk}u^{t}_{k}) = sgn(\sum_{i=1,i\ne k}^d a_{ki}u^{t}_{i})$:
then the update factor in equation \ref{feature_update} is similar to scaling $u^{t}_k$. i.e we could say
$\hat{v}_{k}-v_k=\gamma u^{t}_{k}$ where $\gamma$ is a positive constant.

\item case 2: if $sgn(a_{kk}u^{t}_{k}) \ne sgn(\sum_{i=1,i\ne k}^d a_{ki}u^{t}_{i})$: then,
  \begin{itemize}
      \item case 2a): $|a_{kk}u^{t}_{k}|>|\sum_{i=1,i\ne k}^d a_{ki}u^{t}_{i}|$ then here too the update factor in equation \ref{feature_update} is similar to scaling of $u^{t}_{k}$. i.e. we could say $\hat{v}_{k}-v_k=\gamma u^{t}_{k}$ where $\gamma$ is a positive constant.
      \item case 2b): $|a_{kk}u^{t}_{k}|<|\sum_{i=1,i\ne k}^d a_{ki}u^{t}_{i}|$, then here the update factor is such that it shifts opposite to $sgn(u^{t}_k)$. 
      \item case 2c): $|a_{kk}u^{t}_{k}|=|\sum_{i=1,i\ne k}^d a_{ki}u^{t}_{i}|$, then $\hat{v}_{k}-v_k=m \times K \times 0=0$ implying $\hat{v}^{j^*}_{k}=v^{j^*}_k$, thus no change in the feature $k$ after attack.
      \end{itemize}
\end{itemize}

In particular, when target item true rating distribution possess certain structures, we have the following key observations regarding the relative signs and magnitudes of $a_{kk}u^{t}_{k}$ and $\sum_{i=1,i\ne k}^d a_{ki}u^{t}_{i}$. Further we show that equation \ref{feature_update} is convergent under certain conditions.
 
 \subsubsection*{1) $|a_{kk}u^t_k|$ reduces as the number of true ratings to target item increases: }
If we look at the term $A^{-1}=(\sum_{i \in \mathcal{U}(j^*)}U_iU_i^T+\lambda I)^{-1}$, we can write using Sherman-Morrison-Woodbury identity \footnote{See Definition \ref{sherman_morrison} in Appendix} that for every new true user $\hat{i}$ with feature vector $U_{\hat{i}}$ providing ratings to item $j^*$, the inverse is updated such that $$\hat{A}^{-1}=(A+U_{\hat{i}}U_{\hat{i}}^T)^{-1}=A^{-1}-\frac{A^{-1}U_{\hat{i}}U_{\hat{i}}^TA^{-1}}{1+U_{\hat{i}}A^{-1}U_{\hat{i}}^T}$$ 
From lemma \ref{update_inverse_diagonal} \footnote{See Lemma \ref{update_inverse_diagonal} in Appendix}, the diagonal elements of updated inverse $\hat{A}^{-1}$ is always less than or equal to the diagonal elements of $A^{-1}$. This implies that as more true user provides ratings, diagonal values of $A^{-1}$ decrease. 
i.e., term $|a_{kk}u^t_k|$ reduces as the number of true ratings to the target item increases.

\subsubsection*{2) $sgn(a_{ki}u^{t}_{i})$ relative to $sgn(a_{kk}u^{t}_{k})$: } By lemma \ref{eqn7_behaviour} \footnote{See Lemma \ref{eqn7_behaviour} in Appendix}, we have that if true ratings to item $j^*$ come exclusively from target cluster users, each term $a_{ki}u^{t}_{i}$ in the summation has a sign opposite to $sgn(a_{kk}u^{t}_{k})$. Conversely, if the source of true ratings is not exclusive to the target cluster, then the sign of each term $a_{ki}u^{t}_{i}$ in the summation does not depend exclusively on $sgn(a_{kk}u^{t}_{k})$ and may not all be the same.

\subsubsection*{3) Convergence of $A^{-1}\tilde{U}_t$ }

Suppose item $j^*$ received a large number of true ratings from true users in RS.
Will $V_{j^*}$ prove difficult to change after the attack?  From case $2c$, we can see that magnitude of updates to the item vector after an attack is reduced only if $A^{-1}\tilde{U}_t$ approaches zero vector. 
In particular, we have 
\begin{restatable}[\it{Proof in Appendix}]{Theorem}{convergence}\label{thm:targetratingslemma}
Given $A^{-1}$, let $N$ true ratings be received by target item such that $A^{-1}$ updates to $\hat{A}^{-1}$. Then column vector $\hat{A}^{-1}\tilde{U}_t$ is guaranteed to converge to zero vector as $N$ approaches infinity only if the $N$ ratings come from the target cluster. For increasing true ratings from any non-target cluster, convergence to zero is not guaranteed.
\end{restatable}

Thus if fewer users from the target cluster provide ratings to item $j^*$, then irrespective of the number of ratings from non-target clusters, $V_{j^*}$ may not be immune to changes from attack.
So we expect the effectiveness of the attack to depend on the number and cluster-wise distribution of true ratings to the target item. $V_{j^*}$ is guaranteed to be protected from attack only when the target cluster provides a large number of true ratings to the item $j^*$.

\subsubsection{Impact of Attack on Non-Target Clusters and Non-Target Items:} \label{leakage_effect} From our analysis of equation \ref{Uconst}, we expect that non-target items in a cluster will not be affected after an attack. This is because the target item's $V_{j^*}$ is updated independent of $V_{j}$ for all items $j \ne j^*$, thus changes to $V_{j^*}$ after the attack does not affect any $V_{j}$.  Attack effects leak to non-target item vectors only via changes to $\tilde{U_t}$ as discussed in Section \ref{leakage_items}.  But, since $\tilde{U}$ is kept constant, post-attack changes are only isolated to the target item. 

The predicted rating of target item $j^*$ across all clusters is calculated by $\tilde{U}^TV_{j^*}$. Thus item vector $V_{j^*}$ is common to all the cluster weight vectors in $\tilde{U}$. This implies that changes to the item vector affect all clusters.
For example, suppose true users have $U=[1,-1]$ or $U=[-1,1]$.   i.e. the first set of users like items with $V=[1,0]$ and dislike items with $V=[0,1]$, but the second set of users are the opposite. Then an attack against the first set of true users keeping $U=[1,-1]$ constant to increase the rating of an item with $V=[0,1]$  may be performed by shifting $V=[0,1]$ to $V[1,0]$. Then such a shift, while it increases the rating in users with $U=[1,-1]$, would decrease the rating in group $U=[-1,1]$. 

But based on observations from real-world datasets in Section \ref{Results}, cluster weight vectors are rarely so simple and opposite in preference values. In fact, with higher $d$ dimensional feature space, the cluster preference vectors enjoy correlation. For example, in Table \ref{tab:ml}, features $k=0,3,4,5,6,7$ have the same sign across all clusters. Similar observations can be made for the Goodreads dataset in table \ref{tab:gr} for features $k=0,2,4,6,8,9$. Thus these features of clusters are positively correlated.
In Table \ref{tab:gr}, feature $k=1$ has a negative sign in cluster $2$ but is positive in all non-target clusters. Thus this feature is negatively correlated with corresponding features of clusters $g=0,1,3$. Similarly, many such features exist in both datasets, with opposite signs between clusters.

\begin{table}[H]

\centering
\subfloat[ML]{%
\label{tab:ml}
   \resizebox{6.5cm}{!}
   {
    \begin{tabular}{|l|l|l|l|l|l|l|l|l|l|l|}
    \hline
        g/k & 0 & 1 & 2 & 3 & 4 & 5 & 6 & 7 & 8 & 9 \\ 
        \hline
        0 & 0.54 & 0.06 & 0.03 & 0.06 & 0.63 & -0.43 & 0.29 & -0.25 & 0.45 & -0.01 \\ 
        \hline
        1 & 0.45 & -0.16 & 0.25 & 0.29 & 0.88 & -0.09 & 0.29 & -0.21 & -0.22 & -0.11 \\ \hline
        2 & 0.71 & 0.61 & -0.13 & 0.08 & 0.63 & -0.04 & 0.18 & -0.01 & -0.12 & -0.13 \\ \hline
        3 & 0.30 & -0.08 & -0.46 & 0.20 & 0.81 & -0.36 & 0.60 & -0.04 & -0.17 & -0.34 \\ \hline
    
    \end{tabular}
    }}
\hfill
  \subfloat[GR]{%
  \label{tab:gr} 
  \resizebox{6.5cm}{!}
   {
    \begin{tabular}{|l|l|l|l|l|l|l|l|l|l|l|}
    \hline
        g/k & 0 & 1 & 2 & 3 & 4 & 5 & 6 & 7 & 8 & 9 \\ \hline
        0 & -0.14 & 0.28 & -0.18 & 0.26 & -0.53 & -0.17 & -0.35 & 0.03 & -0.15 & -0.63 \\ \hline
        1 & -0.24 & 0.47 & -0.14 & 0.29 & -0.08 & -0.09 & -0.48 & 0.04 & -0.57 & -0.26 \\ \hline
        2 & -0.11 & -0.22 & -0.47 & 0.42 & -0.53 & -0.19 & -0.47 & 0.26 & -0.30 & -0.10 \\ \hline
        3 & -0.06 & 0.21 & -0.16 & -0.16 & -0.61 & 0.11 & -0.67 & -0.10 & -0.19 & -0.14 \\ \hline

    \end{tabular}
    }}
    \caption{Cluster-Weight Values for $d=10$ features against $|G|=4$ clusters for Movielens (ML) and Goodreads (GR) datasets}
    
\end{table}

We expect that in such correlated clusters, after attack, the effect leaks to non-target clusters as well.
While the relative change in rating increases in the target cluster, for non-target clusters, the relative change in rating may increase or decrease after attack depending on their weight vector $\tilde{U}_g$ and correlation with updated target item vector $\hat{V}_{j^*}$.

\section{Experiment Evaluation Set-up}\label{set-up}

To illustrate our findings discussed in previous sections, we describe the experimental set-up and report the results in Section \ref{Results}. 

\subsection{Datasets}

We evaluate the effectiveness of the attack on the MovieLens
dataset (943 users rating 1682 movies, contains 100000 ratings from 1-5), widely used in literature for evaluating RS under attack, and the Goodreads 10K dataset (53,424 people rating 10,000 books, 5.9M ratings from 1-5). \footnote{Results from MovieLens-1M dataset is included in the appendix and is similar to the results on other datasets, therefore not reported separately}
We take a dense subset of the Goodreads dataset (since MF is computationally expensive for large datasets), obtained
by selecting the top 1000 users who have provided the most ratings and the top 1682 items rated by these users. This provides us with 1000 users and 1682 items.

\subsubsection{ Synthetic User Generation} \label{U_i generation}
We evaluate by generating synthetic users using Movielens and Goodreads as a baseline where, by construction, the setup considered in this section is intentionally tightly controlled so that we can vary one aspect at a time and study its impact.

From the original R matrix of the two datasets considered, calculate $U, V$, and cluster $U$ using k-means to get group membership of all users. 
For users in each group, calculate the empirical distribution of the ratings for each item. Using this mean and variance of the item ratings for a group $g$, generate the required number of users per group by drawing a vector of random ratings for items, i.e., for item $j$ generate a random variable with the probability of each rating being given by the empirical distribution previously calculated for group $g$, item $j$. Notably, we generate ratings so that the proportion of ratings from group $g$ for each item $j$ matches that in the original Movielens and Goodreads dataset. For example, if an item $j$ received ratings from $10\%$ of the users from cluster $g$ in the original dataset and if we generated $1000$ users per cluster, then we ensure that the item $j$ also receives ratings from $10\%$ of users from cluster $g$ ($=100$ ratings) in this generated dataset. For a target item, the number and distribution of ratings per group are completely controlled while leaving the distribution of ratings for other items much the same as before. This also has the advantage that we can easily generate large numbers of new users cleanly and reproducibly. 

For our study, we generate $4$ clusters (results are similar for any number of clusters chosen, but a cluster size of 4 gives a larger sample set of target items with required empirical mean in a target cluster) with latent feature space of dimension $d = 10$ and a number of items of $1682$ for both datasets. Note that there must be at least $250$ users per cluster. This is because too few users cause fewer ratings to be generated per cluster according to our set-up described. Lack of sufficient ratings per cluster causes the centers of the generated clusters to shift from the original centers. This may cause the membership of generated users to change.
So unless mentioned otherwise, for all our simulation studies, we fix the cluster-wise population to be $250$ per cluster. 
The results reported are the average of over $50$ such datasets generated randomly as described using Movielens and Goodreads as the baseline.

\subsection{Threat Model}
\subsubsection{Target Item} \label{target_item}
The target item is given the maximum rating to try to promote it in the system. 
Specifically, in our experiments, we randomly sample an item from a set of items with an empirical mean rating $\le 3$ \footnote{Alternately, to demote an item, we choose a target item with rating $\ge 3$ and fake users assign fake ratings of minimum value $1$. A result showing demotion is attached in the Appendix under additional results. The conclusions and trends are similar to promotion attacks, hence are not discussed separately} and treat it as the target item. The number and distribution of ratings per cluster are completely controlled for the chosen target item, as detailed in Section \ref{U_i generation}. We thus evaluate our results for different target items based on the number of true ratings it has. 

\subsubsection{Filler Items}\label{filler_item}

The filler items are chosen to aid fake users in correctly entering the target cluster.  We rate filler items such that the columns in the $U$ matrix corresponding to the fake users are closer to those of users in the target cluster than those in others. Randomly choosing items may risk the fake user falling into a non-target cluster. So we choose those that are likely to be favored by the users of the target group. Thus we choose so-called distinguisher items \cite{fast_cold_start} as filler items. Specifically, we choose distinguisher items identified in a cluster with i) mean rating very different and higher from other clusters and ii)  the ratings tend to be consistent/reliable, i.e., the variance is small. These 'popular' items in the target cluster can make fake user profiles more realistic and representative of the target cluster user's preferences.
Also, too low a number of filler items makes it difficult for fake users to enter the target cluster correctly. So, we fix the number of filler items to be the minimum number of items rated by true users in the generated data set. The ratings for these filler items are sampled from the Gaussian distribution using the cluster-wise mean rating and a small standard deviation. 

\begin{figure}[H]
\centering
\subfloat[g=0]{ \includegraphics[width=0.20\textwidth]{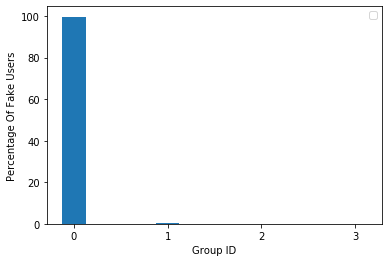}}
\hfill
\subfloat[g=1]{\includegraphics[width=0.20\textwidth]{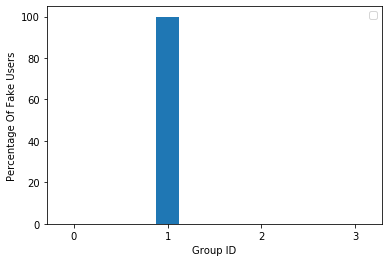}}%
\hfill
\subfloat[g=2]{ \includegraphics[width=0.20\textwidth]{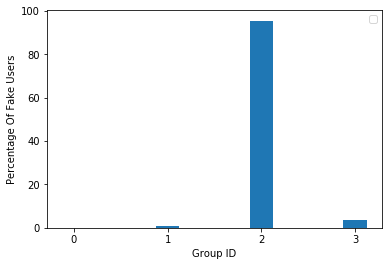}}%
\hfill
\subfloat[g=3]{ \includegraphics[width=0.20\textwidth]{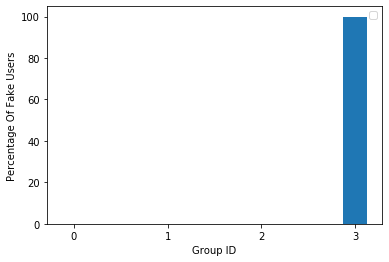}}%
\hfill

\subfloat[g=0]{ \includegraphics[width=0.20\textwidth]{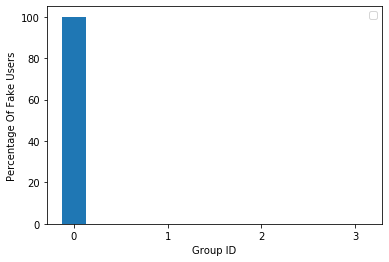}}
\hfill
\subfloat[g=1]{\includegraphics[width=0.20\textwidth]{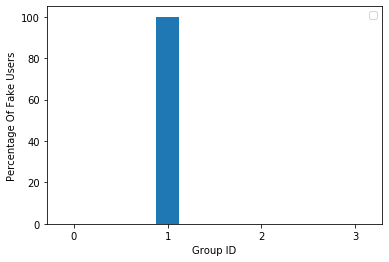}}%
\hfill
\subfloat[g=2]{\includegraphics[width=0.20\textwidth]{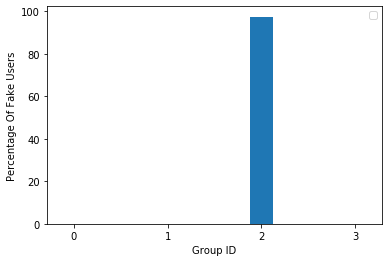}}%
\hfill
\subfloat[g=3]{ \includegraphics[width=0.20\textwidth]{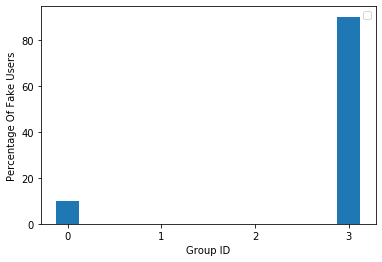}}%
\caption{Plot showing the percentage of fake users entering per cluster when targeting each cluster $g$ for ML and GR Datasets respectively using distinguisher filler items}
\label{8figs}
\end{figure}

In Figure \ref{8figs} we see the percentage of fake users entering the target cluster for our choice of distinguisher filler items. We can observe $>=90\%$ of fake users correctly entering the target cluster. Thus the fake users simulate the preferences of target cluster users accurately.

\subsection{Performance Metric}
We use the change in the rating of the target item relative to the maximum deviation possible as our evaluation metric.

\begin{align} 
\text{Relative Change in Rating}_g = \frac{\mu_f(g,i) - \mu_o(g,i)}{|5-\mu_o(g,i)|}\label{eq:metric}
\end{align}
Where $\mu_f(u, i)$ is the predicted rating of target item $i$ of a user in cluster $g$ after the attack, $\mu_o(g, i)$ is the predicted rating of the target item $i$ of a user in cluster $g$ before the attack and 5 is the maximum rating that can be given to the target item in the datasets that we consider.

We note that it is common in the literature to measure the effectiveness of an attack using either Hit-Rate (HR@N), i.e., the fraction of normal users whose top-N recommendation lists contain the target item, or Prediction Shift (PS), i.e., the difference in the rating of the item before and after the attack indicating by how much the rating has increased after the attack. While the PS metric illustrates whether an attack has the intended effect of increasing rating, it does not measure the attack's power. Thus many studies use HR@N to measure the effectiveness of the attack.

However, in a cluster-based RS, the top $N$ list is common to all users in a cluster. Further, the items in the top $N$ list are RS dependent varying over many factors, such as the threshold rating considered above which it may be put in a recommended list, and the percentage of users who already interacted with that item in the cluster. 
Since our study considers different target items based on the number of true ratings they received, the standard Hit-Ratio is not an ideal evaluation method for us. \footnote{However, we report results for a slightly modified hit-rate definition as an additional result in Appendix}  

Equation (\ref{eq:metric}) gives the shift in predicted rating relative to the maximum deviation possible after an attack.  It therefore directly measures the power of attack on clusters. 

\subsection{Visualising Results}
In each study, we average our results over 50 generated datasets and attack each of these to allow us to study variability in the effectiveness of the attack.  We then use the mean and standard deviation to visualize the relative change in the rating of the target item after the attack. 
We show the results when the adversary targets cluster 2. Results when targeting other clusters are similar and so not reported separately.

\section{Performance Illustration with Data} \label{Results}
Recalculate $U, V$ for the generated user-item ratings in each iteration. This becomes the baseline $U, V$ against which we measure the changes after the attack. Inject fake users and perform update approaches $1$ and $2$ respectively as discussed in Section \ref{update_approaches}. Recall we use a warm start approach when performing updates to $U, V$ and for k-means clustering.
\subsection{Fix $V$, update $\tilde{U}$}\label{result_varyU}

Before we proceed to show the results illustrating our findings, let us briefly recall our key observations of approach $1$ from Section \ref{varyU}:
\begin{itemize}

     \item Relative number of true and fake users plays a role in the effectiveness of the attack. Increasing the number of true users in the target cluster reduces changes to $\tilde{U}_t$ and subsequently the effect of targeted attacks.

    \item  The effect of the attack is isolated to the target cluster when all fake users correctly enter the target cluster. Also, changes to $\tilde{U}_t$ result in 'leakage' of attack to non-target items. The effect of the attack is expected to be pronounced in items highly correlated to the target item.

\end{itemize}

We first illustrate the attack's impact when the target cluster's true user size is increased relative to the fake user size in Section \ref{cluster_pop_effect}. We then proceed to illustrate the leakage effect of attack to non-target items and clusters in Section \ref{leakage_items_res}.

\subsubsection{Varying Ratio of True Users (n) to Fake Users (m) in Target Cluster: }\label{cluster_pop_effect}
 
In this experiment, we compare the relative change in target item predicted rating in the target cluster for varying values of ratio $\frac{n}{m}$. We fix the number of target-item true ratings per cluster throughout the experiment to focus only on changes to $\tilde{U}_t$ due to cluster true user population size. This ensures that any noise due to rating distribution is avoided.
 
To illustrate, we fix a cluster-wise true user population of $250-250-n-250$ (i.e. $250$ users per non-target cluster and $n$ users in target cluster), and a target item rating distribution of $5-5-5-5$ (i.e. $5$ ratings per cluster). Then, we vary the ratio of the number of true users to fake users ($\frac{n}{m}$) in the target cluster. We increase the values from $\frac{n}{m}=0.5$ to $2.5$ for both datasets. Figure \ref{fig:Vconst_vary_Nt} reports the mean and standard standard deviation of change in the target item’s predicted rating when cluster 2 is targeted against the increasing ratio of $\frac{n}{m}$. As expected, it shows decreasing relative change in predicted rating as the ratio increase from $0.5$ to $2.5$ for both Movielens and Goodreads dataset. This is because when $\frac{n}{m}<1$, the presence of a higher number of fake user weights compared to true user weights makes it easier for fake users to shift the cluster weight $\tilde{U}_t$. Thus, we conclude that as the number of true users in the target cluster increases relative to the number of fake users, it becomes harder to attack the item. 
 
\subsubsection{Effect of Attack on Non-Target Items and Non-Target Clusters}\label{leakage_items_res}

\begin{figure}[H]
\centering
\subfloat[Movielens]{
\includegraphics[width=0.35\columnwidth]{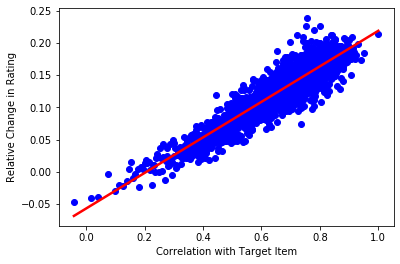}
}
\subfloat[Goodreads]{
\includegraphics[width=0.35\columnwidth]{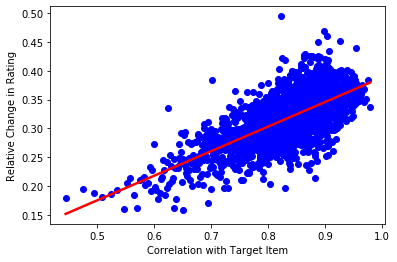}
}\\

\caption{Plot comparing the change in the predicted rating of an item in the target cluster against correlation values between the target item and the other items
}\label{fig:corr_vs_delta}
\end{figure}

Figure \ref{fig:corr_vs_delta}, gives the correlation of each item with the target item v/s the change in the predicted rating of that item in the target cluster. We can see a linear relationship indicating that the effect of the attack on non-target items in the target cluster depends on the correlation between non-target item vector $V_j$ and target item vector $V_{j^*}$. 

\begin{figure}[h]
\centering
\subfloat[Movielens]{
\includegraphics[width=0.35\columnwidth]{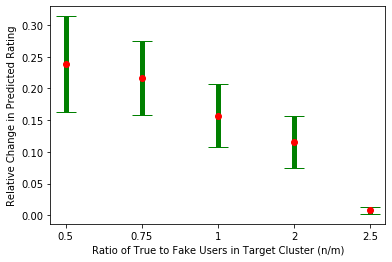}
}
\subfloat[Goodreads]{
\includegraphics[width=0.35\columnwidth]{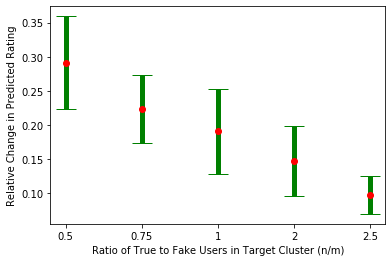}
}\\

\caption{Plot comparing the change in predicted rating in target cluster against the increasing ratio of true users $(n)$ to fake users $(m)$ in the target cluster ($\frac{n}{m}$)}\label{fig:Vconst_vary_Nt}
\end{figure}

\begin{figure}[H]
\centering
\subfloat[Movielens]{
\includegraphics[width=0.35\columnwidth]{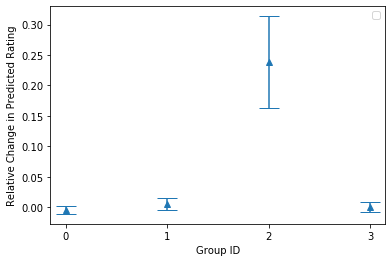}
}
\subfloat[Goodreads]{
\includegraphics[width=0.35\columnwidth]{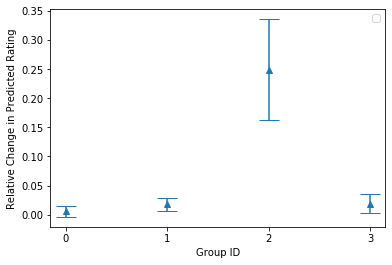}
}\\

\caption{Plot comparing the change in the predicted rating of the target item across clusters when cluster 2 is targeted
}\label{fig:U_delta_versus_clusters}
\end{figure}

The effect of the attack is not expected to be visible in non-target clusters since almost all fake users enter target cluster $t$. In Figure \ref{fig:U_delta_versus_clusters}, we take the case of $\frac{n}{m}=0.5$ and cluster-wise user distribution $250-250-n-250$ which gave the maximum change in predicted rating for the target item in Figure \ref{fig:Vconst_vary_Nt} and show the cluster wise change in the target item predicted rating when cluster 2 is targeted. As predicted, we can see that non-target clusters $0,1,3$ show negligible change in predicted rating after the attack for both datasets. The results are similar for any attack size or cluster population size. Keeping $V$ constant results in the attack effects being isolated to the target cluster alone (when all fake users enter the target cluster). 

\subsection{Fix $\tilde{U}$, Update $V$}\label{result_varyV}

Before we go ahead and show the results illustrating our findings, let us briefly recall our key observations of approach $2$.

\begin{enumerate}
\item  
Increasing true ratings to target items from the target cluster guarantees a reduction of the effect of the attack in the target cluster while increasing true ratings from non-target clusters offers no guaranteed protection from attack.  

\item Attacks against users in a target cluster may 'leak' and affect users in other clusters. The attack may increase or decrease change in predicted rating in other clusters depending on their correlation with the target item feature vector after the attack. This shows that the $V$ vector propagates the effect of the attack to all non-target clusters. This contrasts the case when $V$ is kept constant, causing the attack to be isolated to the cluster where the fake users are present.

\item Simply increasing the true user population ($n$) relative to the fake user population ($m$) does not reduce the change in the target item predicted rating in the target cluster after the attack. This is in contrast to when we update $U$ (keeping $V$ constant), where increasing $n$ relative to $m$ reduces the effect of the attack.
 
\end{enumerate}

First, in Section \ref{0-0-n-0}, we show that increasing true ratings from the target cluster reduce the attack's impact as predicted. Then, we study the attack when the target item is rated by no users in the target cluster but by users in other clusters in Section \ref{n-n-0-n}. For both experiments, we also illustrate the mechanism of how $a_{kk}u^t_k$ and $\sum_{i=1,i\ne k}^d a_{ki}u^{t}_{i}$ of equation \ref{feature_update} work to result in the observed output. Finally, in Section \ref{vary_users_V}, we show that increasing the number of true users in the target cluster does not reduce the impact of the attack on $V_{j^*}$.

\subsubsection{Varying Ratio of True Ratings ($N_t$) to Fake Ratings ($N_f$) in Target Cluster}\label{0-0-n-0}

We consider attacks when the target item has received true ratings in the target cluster but no ratings in the non-targeted clusters. As discussed in previous sections, we expect that increasing true ratings from the target cluster will reduce the attack's impact. Particularly, we study the effect of attack with increasing true ratings $N_t$ from the target cluster given $N_f$ fake ratings. To illustrate, we increase values for $\frac{N_t}{N_f}$ from $0.05 \text{ to } 2.5$ for $N_f=100$ fake ratings and treat it as the target item. We note that the results presented show a similar trend for any choice of increasing ratio.

\subsubsection*{Results:  } Figure \ref{fig:target item ratings} reports mean and standard deviation plots comparing the change in the predicted rating of the target item in target cluster 2 after the attack versus the ratio $\frac{N_t}{N_f}$. It can be seen from both Figures \ref{fig:target item ratings}(a) and \ref{fig:target item ratings}(b) that an item with $\frac{N_t}{N_f}=0.05$  shows the largest mean shift in predicted rating after the attack compared to an item with $\frac{N_t}{N_f}=2.5$ in the target cluster. i.e., it is harder to change the predicted rating of the target item after the attack as the number of true ratings $N_t$ received by the target item increases in the target cluster.

\subsubsection*{Illustration of Update Mechanism: }Let us walk through the update mechanism here. Specifically, in this particular case, since all true ratings come exclusively from the target cluster, as discussed in Section \ref{varyV}, lemma \ref{A_inverseU} case (1) predicts equation \ref{feature_update} to be such that each term in $\sum_{i=1,i\ne k}^d a_{ki}u^{t}_{i}$ always has a sign opposite to $sgn(a_{kk}u^{t}_{k})$. Also, recall that as the true ratings $N_t$ increase, diagonal term $a_{kk}$ decreases.

For $\frac{N_t}{N_f}=0.05$ i.e. when number of true ratings $N_t$ is least, diagonal values $a_{kk}$ are higher such that all features follow mechanism of case $2a: |a_{kk}u^{t}_{k}|>|\sum_{i=1,i\ne k}^d a_{ki}u^{t}_{i}|$, thus updates to feature $v_k$ is similar to scaling of $u^{t}_{k}$. 
Table \ref{tab:dot_product_dist1} reports $\hat{v}_k-v_k$ update values for each feature $k$ of the target item vector for both datasets. Let us focus on $|\hat{v}_{k}-v_{k}|$ corresponding to large $|u^{t}_{k}|$ in both datasets.\footnote{ Recall with $\tilde{U}_t$ constant, change in rating after attack is given by $\tilde{U_t}^T\hat{V}_{j^*}-\tilde{U_t}^TV_{j^*}=\sum_{k=1}^d u^{t}_{k}(\hat{v}^{j^*}_{k}-v^{j^*}_{k})$. Note $\hat{v}^{j^*}_{k}-v^{j^*}_{k}$ corresponding to a large $|u^{t}_{k}|$ will contribute more to change in rating than $\hat{v}^{j^*}_{k}-v^{j^*}_{k}$ corresponding to a lower $|u^{t}_{k}|$ since such low $|u^{t}_{k}|$ will down-weight $\hat{v}^{j^*}_{k}-v^{j^*}_{k}$ values.}  
From the table, we can see that $|\hat{v}_k-v_k|$ corresponding to $k=0,1,4$ in ML and $k=2,3,4,6$ in GR result in large values contributing more to the inner product $\tilde{U}_t^T\hat{V}_{j^*}$ resulting in the higher relative change in rating values for $\frac{N_t}{N_f}=0.05$. 

As the true ratings $N_t$ increases when ratio $\frac{N_t}{N_f}$ increases from $0.05$ to $2.5$, diagonal term $a_{kk}$ decreases, bringing down $|a_{kk}u^{t}_{k}|$ causing the gap between the two terms $|a_{kk}u^{t}_{k}|$  and  $|\sum_{i=1,i\ne k}^d a_{ki}u^{t}_{i}|$ to reduce. Thus $|\hat{v}_{k}-v_k|$  will tend to smaller values. From table \ref{tab:dot_product_dist1}, we note all feature's $|\hat{v}_{k}-v_k|$  decreasing to smaller values for $\frac{N_t}{N_f}=2.5$. 
A similar reduction in magnitude can be observed for the Goodreads dataset. This translates as the effect of attack decreasing for increasing ratio as illustrated by Figure \ref{fig:target item ratings}.

\subsubsection*{Leakage of Attack to Non-Target Clusters: }To see the effect of attack leakage to non-target clusters, Figure \ref{fig:r100_in_gt} further breaks down the change in rating by the user clusters when cluster 2 is targeted in Movielens and Goodreads dataset. It can be seen from Figure \ref{fig:r100_in_gt} that the non-target clusters also show a positive change in the predicted rating. \footnote{We note that there may be target items that, after transformation, result in leakage such that non-target clusters show a negative relative change in rating. But we do not find an item that may illustrate such behavior here. This is because, for our experiments, we choose randomly from a set of target items that has a reliable empirical mean $\le 3$. Such behavior may be found for a larger set of representative target items or in other datasets.} The attack, therefore, effectively fails to be focused on the target cluster. 

\begin{table}
    \centering
    \subfloat[True rating distribution $0-0-N_t-0$]{
    \resizebox{6.5cm}{!}
    {
    \begin{tabular}{|l|l|l|l|l|l|l|l|l|l|l|}
    \hline
        $\frac{N_t}{N_f}$ & k=0 & k=1 & k=2 & k=3 & k=4 & k=5 & k=6 & k=7 & k=8 & k=9 \\ \hline
        \textbf{ML : $\tilde{U}_2$} & \textbf{0.71} & \textbf{0.61} & \textbf{-0.13} & \textbf{0.08} & \textbf{0.63} & \textbf{-0.04} & \textbf{0.18} & \textbf{-0.01} & \textbf{-0.12} & \textbf{-0.13} \\ \hline
        0.05 & 1.15  & 0.99  & -0.23  & 0.15  & 1.03  & -0.07  & 0.29  & -0.02  & -0.20 & -0.20  \\ \hline
        2.5 & 0.28  & 0.23  & -0.05  & 0.04  & 0.28  & -0.05  & 0.07  & -0.03  & -0.05  & -0.06 \\ \hline
        \textbf{GR : $\tilde{U}_2$} & \textbf{-0.11} & \textbf{-0.22} & \textbf{-0.47} & \textbf{0.42} & \textbf{-0.53} & \textbf{-0.19} & \textbf{-0.47} & \textbf{0.26} & \textbf{-0.30} & \textbf{-0.10}  \\ \hline
        0.05 & -0.21  & -0.43  & -0.90  & 0.81  & -1.04  & -0.37  & -0.90  & 0.51  & -0.60 & -0.22  \\ \hline
        2.5 & -0.05  & -0.10  & -0.22  & 0.20  & -0.28  & -0.07  & -0.21  & 0.15  & -0.13  & -0.009 \\ \hline
    \end{tabular}}\label{tab:dot_product_dist1}
    }
     \subfloat[True rating distribution $N_t-N_t-0-N_t$]{
    \resizebox{6.5cm}{!}
    {
    \centering
    \begin{tabular}{|l|l|l|l|l|l|l|l|l|l|l|}
    \hline
        $\frac{N_t}{N_f}$ & k=0 & k=1 & k=2 & k=3 & k=4 & k=5 & k=6 & k=7 & k=8 & k=9 \\ \hline
        \textbf{ML : $\tilde{U}_2$} & \textbf{0.71} & \textbf{0.61} & \textbf{-0.13} & \textbf{0.08} & \textbf{0.63} & \textbf{-0.04} & \textbf{0.18} & \textbf{-0.01} & \textbf{-0.12} & \textbf{-0.13} \\ \hline 
        0.05  & 1.43  & 2.45  & -0.35  & -0.16  & 0.23  & 0.72  & -0.47  & 0.45  & -0.64 & -0.11 \\ \hline
        2.5 & 1.34  & 2.45  & -0.35  & -0.16  & 0.12  & 0.80  & -0.47  & 0.49  & -0.71  & -0.10  \\ \hline
        \textbf{GR : $\tilde{U}_2$} & \textbf{-0.11} & \textbf{-0.22} & \textbf{-0.47} & \textbf{0.42} & \textbf{-0.53} & \textbf{-0.19} & \textbf{-0.47} & \textbf{0.26} & \textbf{-0.30} & \textbf{-0.10} \\ \hline
        0.05 & -0.02  & -1.84  & -1.40  & 1.26  & -0.75  & -0.55  & -0.45  & 1.08  & -0.44 & 0.83  \\ 
        \hline
        2.5 &  -0.01  & -1.90  & -1.37  & 1.21  & -0.56  & -0.49  & -0.40  & 1.09  & -0.47  & 1.06  \\ \hline
    \end{tabular}}\label{tab:dot_product_dist2}
    }
    \caption{Target cluster 2 feature vector and target item update vector $\hat{v^{j^*}_k}-v^{j^*}_k$ values for Movielens and Goodreads Dataset for $d=10$ features}
\end{table}


\subsubsection{ Varying Ratio of True Ratings $(N_t)$ to Fake Ratings $(N_f)$ in Non-Target Clusters}\label{n-n-0-n}

We now consider attacks when the target item has received true ratings in non-target clusters but no ratings in the targeted cluster.   
Particularly, we study the effect of attack with increasing true ratings $N_t$ from each non-target cluster given $N_f$ fake ratings. To illustrate, we increase values of $\frac{N_t}{N_f}$ per non-target cluster from $0.05$ to $2.5$ given $N_f=100$ fake ratings. Since no true ratings come from the target cluster, by theorem \ref{thm:targetratingslemma}, increasing total true ratings may not reduce the attack's impact in the target cluster.

\subsubsection*{Results: }Figure \ref{fig:r5_in_ng} reports the measured mean change in the target item's predicted rating in the targeted cluster against increasing $\frac{N_t}{N_f}$. It can be seen that the decreasing trend is much slower (almost negligible) compared to what was reported in Section \ref{0-0-n-0}, signifying that even with a larger absolute number of true ratings compared to the previous case, the attack is effective in the target cluster. i.e., the large overall number of true ratings fails to protect users in the targeted cluster as presumed. 

\subsubsection*{Illustration of Update Mechanism: }Let us look at the update mechanism here. Since true ratings are not exclusively from the target cluster,  as predicted by lemma \ref{eqn7_behaviour} case (2), equation \ref{feature_update} is such that terms $a_{ki}u^t_i$ in summation do not have the same signs. This reduces the overall magnitude of term $\sum_{i=1,i\ne k}^d a_{ki}u^{t}_{i}$ compared to last section where all the terms in the summation had the same sign. Thus even though diagonal terms $a_{kk}$ show a decreasing trend with increasing $N_t$, the behavior of summation terms may cause $|a_{kk}u^{t}_{k}|>|\sum_{i=1,i\ne k}^d a_{ki}u^{t}_{i}|$ even at higher ratios for some features, resulting in the reported greater relative change in predicted rating at these ratios. For example, note in table \ref{tab:dot_product_dist2} how feature $v^{j^*}_k$ especially corresponding to larger magnitude $u^{t}_{k}$ ($k=0,1,4$ in ML and $k=2,3,4,6$ in GR), result in large magnitude of $\hat{v}^{j^*}_k$ despite increasing ratio. This contrasts our observation from table \ref{tab:dot_product_dist1}. 

\subsubsection*{Leakage of Attack to Non-Target Clusters: }
\begin{figure}
\centering
\subfloat[Movielens]{
\includegraphics[width=0.35\columnwidth]{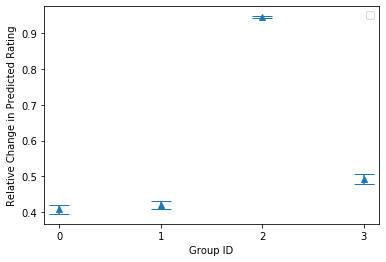}
}
\subfloat[Goodreads]{
\includegraphics[width=0.35\columnwidth]{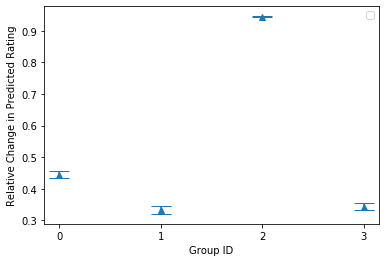}
}\\

\caption{Plot comparing the relative change in predicted rating of target item across clusters when targeting cluster 2 for $\frac{N_t}{N_f}=0.05$ 
}\label{fig:r100_in_gt}
\end{figure}

\begin{figure}
\centering
\subfloat[Movielens]{
\includegraphics[width=0.35\columnwidth]{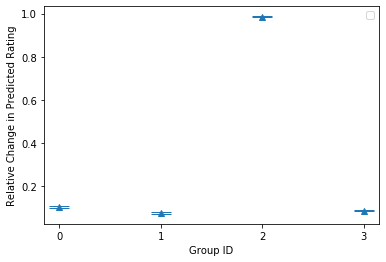}
}
\subfloat[Goodreads]{
\includegraphics[width=0.35\columnwidth]{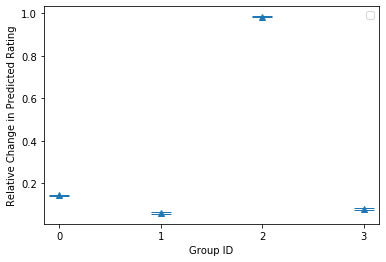}
}
\caption{Plot comparing the relative change in predicted rating of target item across clusters against ratio $\frac{N_t}{N_f}=0.05$ per non-target cluster.}
\label{fig:r100_in_ng}
\end{figure}

Similar to in Section \ref{0-0-n-0}, we expect to see leakage effect in non-target clusters.  This is illustrated by Figure \ref{fig:r100_in_ng} which breaks down the change in rating by clusters when cluster 2 is targeted in Movielens and Goodreads datasets. You can see that the attack's impact on non-targeted clusters is lower here than suggested by Figure \ref{fig:r100_in_gt}. This is because compared to the last section where all positively correlated features across clusters underwent case $2a$ mechanism, due to properties of $A^{-1}$ here, $\hat{v^{j^*}_k}-v^{j^*}_k$ corresponding to many positively correlated cluster features (e.g.: $k=3,5,6,7$ in ML and $k=9$ in GR) undergo case $2b$ mechanism causing a shift in directions dissimilar to not only $u^{t}_{k}$ but all corresponding  $k^{th}$ features across non-target clusters $g \in \mathcal{G}\setminus {t}$.
 
Thus in these two datasets, the properties of $U, V $ values are such that unless true ratings also come from the target cluster, it is tough to reduce the effect of attack within the target cluster.

\begin{figure}
\centering
\subfloat[Movielens]{
\includegraphics[width=0.35\columnwidth]{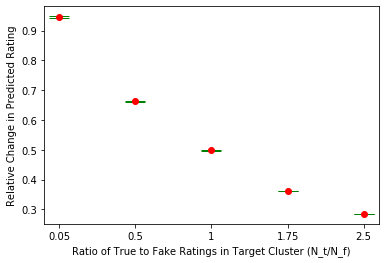}
}
\subfloat[Goodreads]{
\includegraphics[width=0.35\columnwidth]{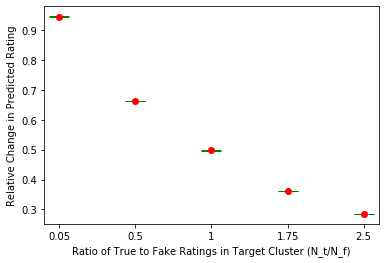}
}\\

\caption{Plot comparing the relative change in the predicted rating of target item against increasing ratio $\frac{N_t}{N_f}$ in the target cluster 
}\label{fig:target item ratings}
\end{figure}

\begin{figure}
\centering
\subfloat[Movielens]{
\includegraphics[width=0.35\columnwidth]{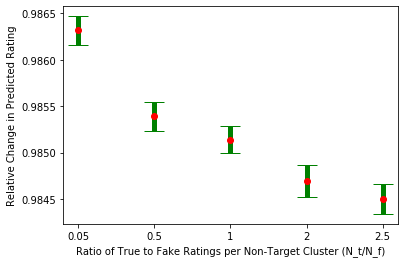}
}
\subfloat[Goodreads]{
\includegraphics[width=0.35\columnwidth]{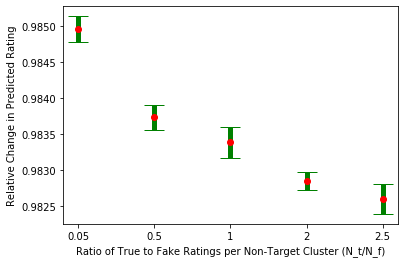}
}
\caption{Plot comparing the relative change in the predicted rating of target item against increasing ratio $\frac{N_t}{N_f}$ per non-target cluster}
\label{fig:r5_in_ng}
\end{figure}

\begin{figure}
\centering
\subfloat[Movielens]{
\includegraphics[width=0.35\columnwidth]{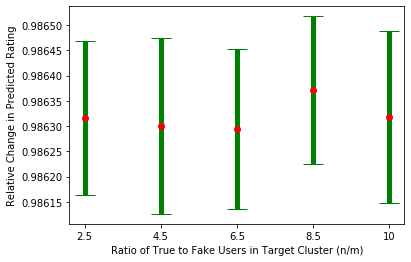}
}
\subfloat[Goodreads]{
\includegraphics[width=0.35\columnwidth]{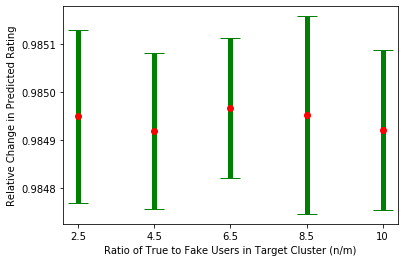}
}\\

\caption{Plot comparing the relative change in the predicted rating of target item against increasing target cluster size $n$ 
}\label{fig:size_vary_V}
\end{figure}

\subsubsection{Varying Ratio of True Users (n) to Fake Users (m) in Target Cluster}\label{vary_users_V}

In this section, we investigate the impact of the target cluster's true user size $(n)$ relative to the fake user size $(m)$ on the effectiveness of the attack. 

 For this experiment, we fix the distribution of ratings as $\frac{N_t}{N_f}=0.05$ and $N_f=m=100$ similar to in Section \ref{0-0-n-0} since it resulted in the maximum change in relative rating in Figure \ref{fig:target item ratings}. Then we increase the ratio of $\frac{n}{m}$ from $2.5$ to $10$. Figure \ref{fig:size_vary_V} reports the mean and standard deviation of change in the target item’s predicted rating when cluster 2 is targeted against increasing values of $\frac{n}{m}$. We can see how the increasing target cluster size does not affect the attack's impact. The relative change in predicted rating after the attack remains almost the same as expected.

\section{Other Feature Based Recommendation Models}
Previously, our approach involved using a dot product to merge the user and item embeddings. However, there has been a shift towards leveraging neural networks to predict ratings. We're exploring if the insights from this study apply to other feature-based models as well.
\subsection{Learning a Dot Product Model with MLP}
In this scenario, we replicate our experiments using a Neural Collaborative Filtering (NCF) based RS. Typically, a Multilayer Perceptron (MLP) serves as the network architecture here \cite{NCF}.
We utilize the user and item vectors, $U$ and $V$, obtained through MF. However, instead of employing a dot product, we concatenate these vectors and feed them into MLP, where the final predicted rating is the output. We employ two hidden layers for MLP. Given the latent factors size  $d$, we use an input layer of size $2d$ consisting of the concatenation of the two embeddings and $2$ hidden layers of sizes [2d, d]. Also, we use the ReLU as the activation function and Adam optimiser. We repeat the experiments and provide the results below for this set-up in ML-100k dataset \footnote{The results are similar in the other datasets, thus are not shown separately.}.

\begin{figure}
\centering
\subfloat[]{
\includegraphics[width=0.35\columnwidth]{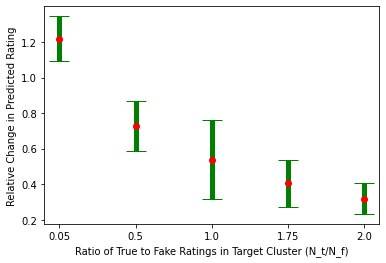}
}
\subfloat[]{
\includegraphics[width=0.35\columnwidth]{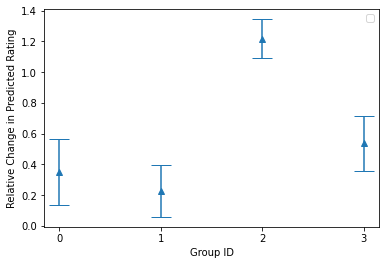}
}\\

\caption{MLP based RS: Plot comparing the relative change in predicted rating of target item when updating only $V$ after the attack a) against increasing ratio $\frac{N_t}{N_f}$ in target cluster  b) across user clusters for $\frac{N_t}{N_f}=0.05$ in the target cluster
}\label{fig:mlp_v}
\end{figure}

In Figure \ref{fig:mlp_v} (a), we demonstrate the change in predicted rating as the ratio $\frac{N_t}{N_f}$ increases. In this illustration, we maintain the user vector $U$ constant while updating the item vector $V$. These vectors are then fed into a MLP for rating prediction. As observed in our previous findings, the impact of attacks through shifts to $V$ is evident across clusters in Figure \ref{fig:mlp_v} (b), with the effect diminishing in the target cluster as the number of true ratings in the target cluster increases.

\begin{figure}
\centering
\subfloat[]{
\includegraphics[width=0.35\columnwidth]{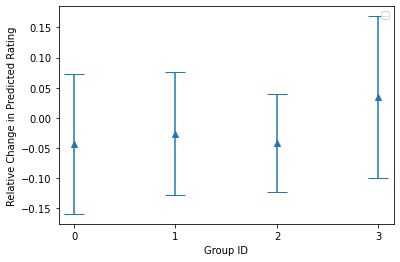}
}

\caption{MLP based RS: Plot comparing the relative change in predicted rating of target item when updating only $U$ after attack across user clusters for $\frac{N_t}{N_f}=0.05$ and $\frac{n}{m}=0.5$ in target cluster
}\label{fig:mlp_u}
\end{figure}

Conversely, when $V$ is kept constant and $U$ is updated, inputting them into the MLP results in minimal changes after the attack across user clusters as illustrated in Figure \ref{fig:mlp_u}. Regardless of the ratios $\frac{N_t}{N_f}$ and $\frac{n}{m}$, the observed effect remains negligible. 

These results are in agreement with our observation that it is item feature $V_{j^*}$ of targeted item $j^*$ that propagates the effect of the attack in the RS.

\subsection{Dot Products at the Output Layer of DNNs}

We expect the results to carry over to any approach that uses a dot product to predict the ratings.
Consider the configuration where we employ a deep neural network architecture to learn the user and item feature matrices, as in \cite{deep_MF}. Specifically, we utilize two multi-layer networks (comprising 3 hidden layers and an output layer of size $d=10$) to extract latent user and item feature matrices of dimension $d$. The predicted rating is then calculated using the dot product of the obtained embeddings/feature matrices, with the sparse ratings matrix serving as input.

\begin{figure}
\centering
\subfloat[]{
\includegraphics[width=0.35\columnwidth]{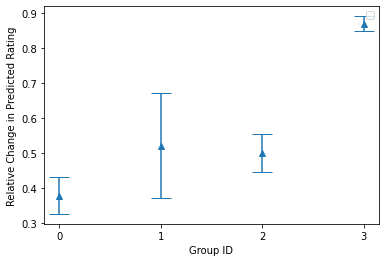}
}
\subfloat[]{
\includegraphics[width=0.35\columnwidth]{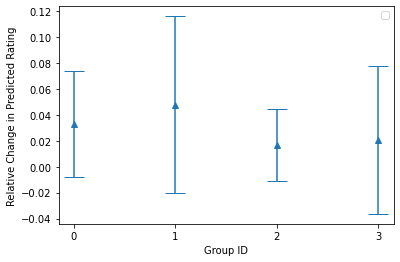}
}\\

\caption{DMF based RS: Plot comparing the relative change in the predicted rating of target item across user clusters when a) update $V$, keep $U$ constant b) update $U$, keep $V$ constant when targeting cluster 2 
}\label{fig:DMF}
\end{figure}

Following the same procedure, Figure \ref{fig:DMF} illustrates results on ML-100k dataset \footnote{The results are similar in the other datasets, thus are not shown separately.}. Consider the overall ratio between true to fake ratings as $0.05$ indicating a target item with very few ratings in the RS. When retraining the neural network after an attack with $100$ fake ratings and maintaining $U$ constant while updating $V$, the effects of the attack permeate across user clusters as observed in Figure \ref{fig:DMF}(a), similar to the observations in the previous cases. Conversely, if $V$ remains unaltered after the attack, the impact is negligible in comparison, as shown in Figure \ref{fig:DMF}(b).

The extent of the effect across clusters depends on the correlation between the $U_i$ values of users and $V_{j^*}$ value of targeted item $j^*$. Though not reported separately, the effect of attack on clusters due to updates to $V$ also reduces with increasing true ratings to the target item, similar to our previous observations.  

\section{Discussion}

The objective of the targeted attack study on MF-based RS is to gain a better understanding of how attacks propagate and how the individual behaviors of feature matrices $U$ and $V$ contribute to this propagation. By holding one feature matrix constant and updating the other, we isolated the effects of attacks on each matrix and observed how these effects translate to the targeted promotion of items in RS. 

We concluded how $U_i$ values corresponding to all true users $i$ in a RS are unaffected by the injection of fake ratings when only matrix $U$ is updated to incorporate the new ratings, keeping $V$ constant. The group feature vector $\tilde{U}_t$ of the target group $t$ is affected due to the presence of fake users in the target group after the attack. This effect  on $\tilde{U_t}$ decreases when the number of true users in the target cluster increases. Compared to the change in rating due to changes in item vector $V$, the effect in predicted ratings due to changes in $\tilde{U_t}$ is negligible i.e. we find that the target item vector $V_{j^*}$ is more sensitive to the injection of fake ratings. It is the compromised $V_{j^*}$ that leaks attack effects to all the clusters (even when the clusters do not contain the presence of fake users).

For our analysis, we examined a single update step for either the $U$ or $V$ matrix when new users join the system. We anticipate that the analysis can be extended during the model's alternating updates until convergence. As we discussed in Section \ref{varyU}, $U_i$ relies on the item vector $V_j$ and the rating value $R_{i,j}$ of all items $j \in \mathcal{V}(i)$ rated by user $i$. In the scenario of a single update, the $U_i$ values for genuine users remain constant since $V$ is fixed. However, when we continue alternating updates until convergence, subsequent updates to $U_i$ involve the most recently updated $V$, potentially leading to changes in $U_i$ for true users. These changes are expected to be modest given that $U_i$ depends on all items rated by user $i$, and since the target item $j^*$ is just one item in the set, the alterations may not be substantial.

However, the update of $V_{j^*}$ for a target item $j^*$ can result in significant alterations in response to fake ratings because $V_{j^*}$ relies on all the ratings received by item $j^*$ as discussed in Section \ref{varyV}. Consequently, after convergence, both $U_i$ and $V_{j^*}$ are expected to undergo modification, with $V_{j^*}$ experiencing the most notable changes comparatively. The findings from the experimental results align with our earlier observations and are detailed in Section \ref{standardMF} in the Appendix.

We expect that the results discussed, provide insight into defense approaches likely to be effective against RS.
\subsection{Possible Defence Mechanisms}
To protect the columns of $V$ matrix from possible attacks, increasing true ratings to items provides a way. Therefore, providing dummy ratings to items rated less frequently in a cluster might provide immunity against malicious changes to its item vector. However, we observed that guaranteeing a reduction in the impact of the attack is not always certain, as, only increasing true ratings from the target cluster is guaranteed to achieve this compared to increasing ratings from non-target clusters. Alternately, updates to latent feature matrices need not be performed frequently together. By updating matrix $V$ less frequently and updating only the $U$ matrix when new users enter, we can decelerate the propagation of attack effects across user clusters and allow monitoring for suspicious trends in item ratings or feature vectors over time. 

The success of the attack depends on how much it can shift the item vector distribution significantly. We propose a detection approach that examines the shift in item vectors \cite{ivd}. If the feature vector of an item changes significantly after a block of ratings enters the RS, we can suspect that the ratings are fake and avoid using them to train the $U, V$ matrices.

\begin{figure}[H]
\centering
\subfloat[Effect of Increasing True Ratings on $V_{j^*}$]{
\includegraphics[width=0.50\columnwidth]{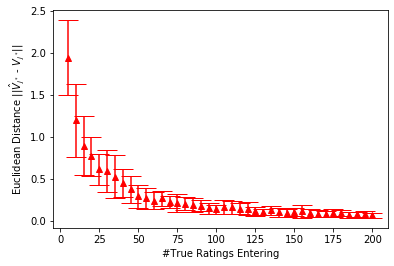}
}
\subfloat[$V_{j^*}$ Deviation from True,Fake Ratings]{
\includegraphics[width=0.50\columnwidth]{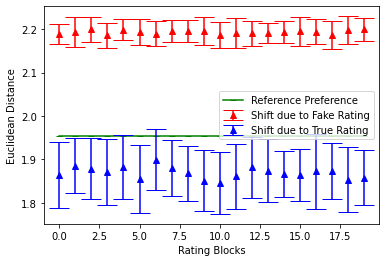}
}
\caption{Plot illustrating deviation of Item Vector in MovieLens 100k dataset}
\label{V}
\end{figure}

Let's consider a scenario where the target item $j^*$ initially lacks any ratings in the RS. We randomly choose a group and a user from that group to provide a rating. The iterative changes in distance (mean and standard deviation) between the updated $\hat{V}_{j^*}$ and the last updated $V_{j^*}$ after each block of true ratings are visualized in Figure \ref{V}(a). The observation reveals that the item vector undergoes fewer shifts from its previous state after approximately $100$ total true ratings. This suggests that the item's preference has been established, and additional true ratings contribute little new preference information.  
 
Given the target item $j^*$, we randomly select a cluster $g$ as a reference in the d-dimensional vector space. We hypothesize that further real ratings will not significantly shift the vector $V_{j^*}$ in reference to cluster $g$. This stems from the notion that true ratings seldom cause substantial changes in current preferences, as previously illustrated. We raise suspicions of fraudulent ratings if any block of new ratings results in a significant deviation of the item vector from the reference cluster $g$. Such deviations may suggest an attempt to manipulate recommendations by shifting the item vector into a different region of the vector space.

To demonstrate this concept, Figure \ref{V}(b) computes the distance ($D_{g,\hat{j}^*}$) between $\hat{V}_{{j^*}}$ to cluster $g$  and compares it to the distance ($D_{g,j^*}$) between initial $V_{j^*}$ and cluster $g$  for each false (indicated by red plot) 
and true rating blocks (indicated by blue plot). As we can see,  each block of fake ratings \footnote{We select a block of $20$ new ratings because attack sizes less than $1\%$ do not lead to significant changes in the item vector, as discussed in \cite{ivd}} (modelled as an Average Attack) leads to a more pronounced deviation from the initial threshold $D_{g,j^*}$ representing reference preference information (depicted in the green plot). This shift suggests a change in preferences. Conversely, each block of real ratings results in shifts that are relatively close to the reference preference $D_{g,j^*}$.

This observation allows us to recognise and eliminate fake ratings from the RS \footnote{This defense mechanism is detailed in \cite{ivd}}. Note that the model can continue to recommend to fake users. Thus, only suspected ratings are removed before the RS retrains the $U, V$ to generate an updated prediction matrix.

\section{Conclusion}

This paper studied the effect of user-cluster targeted data poisoning attacks on an MF-based RS by evaluating the changes after the attack on user and item feature matrices $U, V$. We analysed the mechanism of how $U$ and $V$ matrices change after the injection of fake ratings and how these changes help propagate targeted attacks. We also illustrated our findings using two real-world datasets. We further showed that the effectiveness of an attack on a target item's feature vector is influenced by the cluster-wise distribution of the number of ratings received by the target item. Our findings indicate that items with fewer ratings within the target cluster are more vulnerable to attacks. The number of ratings they receive in other clusters does not necessarily ensure protection from attacks within the target cluster. We conclude that a simple attack with limited knowledge of user preferences suffices to target a specific user cluster precisely. 

For our future work, we intend to use these observations to research different defensive approaches in MF-based RS that are simple yet effective and can be easily used in existing systems. 


\begin{acks}
This work was supported by Science Foundation Ireland (SFI) under grant 16/IA/4610
\end{acks}

\bibliographystyle{ACM-Reference-Format}
\bibliography{ref}

\appendix
\section{Standard Properties Of Positive Definite/ Semi-Definite Matrices }\label{PSD_prop}

\begin{Definition}\label{PSD}
A matrix $M$ is called Positive Semi-Definite (PSD) if it is symmetric and its eigen values are non-negative. If the eigen values are positive, they are called Positive Definite (PD) Matrices. Equivalently, we can say that if the matrix $M$ is symmetric and satisfies $x^TMx>=0$ for a non-zero vector $x$, then $M$ is PSD and if $M$ is symmetric and satisfies $x^TMx>0$ for a non-zero vector $x$, then $M$ is PD. 
\end{Definition}

\begin{Lemma}\label{PSD_ProductofTranspose}
Any $n \times n$ matrix of the form $M=BB^T$ is PSD for any $n \times m$ matrix $B$.
\end{Lemma}
\begin{Proof}\setlength\hangindent{10pt}
By definition \ref{PSD}, for matrix $M$ to be PSD, we require that it be
\begin{itemize}
    \item Symmetric i.e. $M^T=M$
    \item $x^TMx>=0$ for any non-zero vector $x \in \mathbb{R}^n$
\end{itemize}

We have $M^T=(BB^T)^T=BB^T=M$ by the property of transpose $(AB)^T=B^TA^T$ and

Also $x^TMx=x^TBB^Tx=x^TB(x^TB)^T$. Define $y=x^TB$ such that
$$x^TMx=x^TB(x^TB)^T=yy^T=\sum_{i=1}^{m} y_i^2 \ge 0$$

Thus $M=BB^T$ is a Positive Semi-Definite Matrix.
\end{Proof}

\begin{Lemma}\label{PSD_sum}
Let $A$ be a PSD matrix, and $B$ be a PD matrix. Then their sum $A+B$ is a PD Matrix.
\end{Lemma}
\begin{Proof}
By definition \ref{PSD}, for any non-zero vector $x$, the quadratic form $xAx^T>=0$ if $A$ is a PSD matrix and $xAx^T>0$ if $A$ is a PD matrix. 
So a matrix defined as the sum of a PSD matrix $A$ and a PD matrix $B$ satisfies $$x(A+B)x^T=xAx^T+xBx^T>0$$ making the resulting matrix PD.   
\end{Proof}

\begin{Lemma}\label{PSD_inverse}
Let $M$ be a PSD matrix, then its inverse $M^{-1}$ is also a PSD.
\end{Lemma}
\begin{Proof}
 By definition \ref{PSD}, given matrix $M$ is PD, then it satisfies the following properties
\begin{itemize}
    \item Symmetric i.e. $M^T=M$
    \item and has positive eigenvalues
\end{itemize}

We can write $(M^{-1})^T=M^{-1}$ (since $M$ is invertible and satisfies property $(M^{-1})^T=(M^T)^{-1}$).

Also, we know that eigenvalues of $M^{-1}$ are inverse of the eigen values of $M$. Since, by definition, eigenvalues of $M$ are positive, eigen values of $M^{-1}$ are also positive. 
Therefore, $M^{-1}$ is also PSD.
\end{Proof}

\begin{Lemma}\label{PSD_diagonalpositive}
 If the matrix $M$ is PD, then all diagonal elements of $M$ are positive.   
\end{Lemma}
\begin{Proof}
The proof follows by considering the property $x^TMx>0$ for any PD Matrix $M$. If we set $x$ to be the standard basis vector with $x_k = 1$ and $x_i = 0$ for all $i \ne k$, then $x^TMx = M_{kk}>0$, which means that all the entries in the diagonal of $M$ are positive.    
\end{Proof}

\section{Sherman-Morrison Formula}
Given $M$ is a square $n \times n$ matrix whose inverse matrix we know, for the simple case of a rank-1 perturbation to $M$, Sherman–Morrison formula provides a method to find the updated rank-1 change to the inverse.

\begin{Definition}\label{sherman_morrison}
 Given $M$ is a square $n \times n$ matrix whose inverse matrix we know, $u$ and $v$ are $n \times 1$ column vectors defining the perturbation to matrix $M$ such that $u_iv_j^T$ is added to $M_{i,j}$, then we can find the inverse of the modified $M$ by Sherman-Morrison formula

 $$(M+uv^T)^{-1}=M^{-1}-\frac{M^{-1}uv^TM^{-1}}{1+v^TM^{-1}u}$$
\end{Definition}
\noindent For a generalised rank-$k$ perturbation to $M$, the updated inverse is given by the following formula.
\begin{Definition}\label{woodbury_identity}
 Given $M$ is a square $n \times n$ matrix whose inverse matrix we know, $U$ and $V$ are $n \times k$ matrices defining rank $k$ perturbation to matrix $M$, then we can find the inverse of the modified $M$ 
 
 $$(M+UV^T)^{-1}=M^{-1}-M^{-1}U(I_k+V^TM^{-1}U)^{-1}V^TM^{-1}$$
\end{Definition}

\begin{Lemma}\label{update_inverse_diagonal}
 if $M^{-1}$ is a PSD matrix and $\hat{M}=M+uu^T$ for non-zero $u \in \mathbb{R}^{n \times 1}$, then diagonal elements of $\hat{M}^{-1}$ is less than or equal to the diagonal elements of matrix $M^{-1}$. 
\end{Lemma}

\begin{Proof}
 We can find the inverse of $\hat{M}$ by Sherman-Morrison formula

 $$(M+uu^T)^{-1}=M^{-1}-\frac{M^{-1}uu^TM^{-1}}{1+u^TM^{-1}u}$$

 From the expression we note that the numerator of the subtracting term $(M^{-1}u)(u^TM^{-1})$ is the product of a matrix and its transpose $((M^{-1}u)(M^{-1}u)^T)$, so $M^{-1}uu^TM^{-1}$ is PSD matrix by lemma \ref{PSD_ProductofTranspose} and therefore has positive diagonal values by lemma \ref{PSD_diagonalpositive}.
 Since $M^{-1}$ is a PSD matrix and the quadratic form $u^TM^{-1}u>=0$ for any non-zero vector $u$, the denominator of subtracting term is always positive. So we can say that the diagonal elements of the subtracting term are $>=0$. Then we can conclude that the diagonal elements of updated inverse is less than or equal to the diagonal elements of matrix $M^{-1}$.
\end{Proof}


\section{Proofs}

\subsection{Deriving Recursive Updates to $V_j$}\label{derivingV}
\subsubsection{Proof of Theorem \ref{V_iterative}}
\begin{Proof}
For simplicity, let $\Lambda$ gather together the feature vectors $U_i$ of all users $i$ who rated target item $j^*$ such that $\Lambda\Lambda^T=\sum_{i \in \mathcal{U}(j^*)}U_iU_i^T$ and let $r$ be a column vector of ratings received by the target item from these true users. 
We can write equation \ref{Uconst} as,
$$V_{j^*}=(\Lambda\Lambda^T+\lambda I)^{-1}\Lambda r=A^{-1}\Lambda r$$  where $A=\Lambda\Lambda^T+\lambda I$. After the block of fake users enter, we write $\hat{U}=[\Lambda,X]$ and $\hat{U}\hat{U}^T=\Lambda\Lambda^T+XX^T$. Also $\hat{r}=\begin{bmatrix}
r  \\
y  
\end{bmatrix}
$
and
$\hat{A}^{-1}=(\hat{U}\hat{U}^T+\lambda I)^{-1}=(A+XX^T)^{-1}$. We now calculate the updated $\hat{V_{j^*}}$. We know that, 
\begin{align*}
 \hat{V_{j^*}}&= \hat{A}^{-1}\hat{U}\hat{r} = \hat{A}^{-1}\left(\Lambda r+Xy\right)= \left(A+XX^T\right)^{-1}\left(\Lambda r+Xy\right) \\
        &\text{Applying Sherman Morrison/Woodbury formula to the inverse (see Definition \ref{sherman_morrison})}\\
        &= \left(A^{-1}-A^{-1}X\left(I+X^TA^{-1}X\right)^{-1}X^TA^{-1}\right) \left(\Lambda r+Xy\right)\\
        &= A^{-1}\Lambda r-A^{-1}X\left(I+X^TA^{-1}X\right)^{-1}X^TA^{-1}\Lambda r+A^{-1}Xy-A^{-1}X\left(I+X^TA^{-1}X\right)^{-1}X^TA^{-1}Xy \\
        &\text{We have, } V_{j^*}=A^{-1}\Lambda r \text{. substituting for $A^{-1}\Lambda r$ and rearranging, we get, }\\
        &= V_{j^*}-A^{-1}X\left(I+X^TA^{-1}X\right)^{-1}X^TV_{j^*}+A^{-1}X\left(I- \left(I+X^TA^{-1}X\right)^{-1}X^TA^{-1}X\right)y \\
        &\text{Substituting } I=(I+X^TA^{-1}X)^{-1}(I+X^TA^{-1}X) \text{ inside third term, we get,}\\
        \begin{split}
        &= V_{j^*}-A^{-1}X\left(I+X^TA^{-1}X\right)^{-1}X^TV_{j^*}+ \\ &A^{-1}X\left(\left(I+X^TA^{-1}X\right)^{-1}\left(I+X^TA^{-1}X\right)-\left(I+X^TA^{-1}X\right)^{-1}X^TA^{-1}X\right)y 
        \end{split}\\
        &= V_{j^*}-A^{-1}X\left(I+X^TA^{-1}X\right)^{-1}X^TV_{j^*}
        +A^{-1}X\left(\left(I+X^TA^{-1}X\right)^{-1}\left(I+X^TA^{-1}X-X^TA^{-1}X\right)\right)y\\
       &=V_{j^*}-A^{-1}X\left(I+X^TA^{-1}X\right)^{-1}X^TV_{j^*}+A^{-1}X\left(I+X^TA^{-1}X\right)^{-1}y\\
       &=V_{j^*}+A^{-1}X\left(I+X^TA^{-1}X\right)^{-1}\left(y-X^TV_{j^*}\right)
\end{align*}
\end{Proof}

Finally, we have,
\begin{align} \label{Uconst_iter}
\hat{V_{j^*}}-V_{j^*}&=A^{-1}X\left(I+X^TA^{-1}X\right)^{-1}\left(y-X^TV_{j^*}\right)
\end{align}
Here we have three terms $a=A^{-1}X$, $b=\left(I+X^TA^{-1}X\right)^{-1}$ and $c=\left(y-X^TV_{j^*}\right)$. Using Lemma $6,7,8,9,10$ (next section) we re-write equation \ref{Uconst_iter}  as
\begin{align*} 
\hat{V_{j*}}-V_{j*} &= m \times K \times \left(A^{-1}\tilde{U}_t\right) 
\end{align*}
where $K=bc$ and $m, K$ are positive constants.

\subsection{Re-writing Equation \ref{Uconst_iter}}
If we look at the term $A^{-1}=(\Lambda\Lambda^T+\lambda I)^{-1}$ in equation \ref{Uconst_iter}, since $\Lambda\Lambda^T$ is a multiplication of a matrix with its transpose, it is positive semi-definite (PSD) matrix, and $\lambda I$ is a positive definite (PD) matrix. \footnote{See Definition \ref{PSD} and Lemma \ref{PSD_ProductofTranspose} in Appendix \ref{PSD_prop}} Then $(\Lambda\Lambda^T+\lambda I)$ is a PD matrix \footnote{See Lemma \ref{PSD_sum} in Appendix \ref{PSD_prop}}. Since the inverse of a PD matrix is also PD, $A^{-1}$ is a PD matrix. \footnote{See Lemma \ref{PSD_inverse} in Appendix \ref{PSD_prop}} Because $A^{-1}$ is a PD matrix and since columns of $X$ are identical to $\tilde{U}_t$, by lemma \ref{termb} we can say that term $X^TA^{-1}X$ in $b$ results in an $\mathbb{R}^{n \times n}$ matrix with identical values for all the elements.

\begin{theoremEnd}{Lemma}
\label{termb}
    Let $M=X^TA^{-1}X$ and $x_i,x_j\in \mathbb{R}^{d \times 1}$ be the $i^{th}$ and $j^{th}$ column in $X$. Given $x_i=x_j=\tilde{U}_t$ and $A^{-1}$ is a PD matrix, then $(m)_{ij}=x_i^TA^{-1}x_j$ is a positive constant.
\end{theoremEnd}

\begin{Proof}
 \begin{align*}
    (m)_{ij}=x_i^TA^{-1}x_j &= \tilde{U}_t^TA^{-1}\tilde{U}_t
                      \end{align*}
 We know by Definition \ref{PSD} that pre-multiplying and post-multiplying a PD matrix by the same vector takes a quadratic form and always results in a positive number provided the multiplying vector is non-zero.
\end{Proof}
So $X^TA^{-1}X$ results in a matrix with all elements having identical values. By lemma \ref{b_inverse} the inverse of $I+X^TA^{-1}X$ results in a matrix with all diagonal elements having identical values and the non-diagonal elements having identical values. 
\begin{Lemma}\label{b_inverse}
For any matrix $M \in \mathbb{R}^{n \times n}$ with identical elements, the inverse of $I+M$ results in a matrix with all diagonal elements having identical values and the non-diagonal elements having identical values.     
\end{Lemma}
\begin{Proof}
For any $M \in \mathbb{R}^{n \times n}$ with identical elements, we can write $\left(I+M\right)^{-1}= \left(I+ec^T\right)$, for $M=ec^T$ where $e=[1,1,1,\cdots,1]^T$ 
and $c$ is a column of $M$. Applying Sherman-Morrison formula, we get 
$$\left(I+M\right)^{-1}=I-\frac{M}{1+c^Te}$$ Note that the denominator is a positive constant. So we can say that the resulting matrix will have all diagonal elements the same and all non-diagonal values the same.
\end{Proof}

Now we have term $b$. Next, we proceed to evaluate the term $c$.
Term $c=y-X^TV_{j^*}$ says how much of target item fake rating vector $y$ with all its value equal to the highest rating $5$ can be explained by the existing weight vector $V_{j^*}$. If $V_{j^*}$ before the attack can explain the vector $y$ then term $c$ will be a zero vector and equation \ref{Uconst_iter} results in $\hat{V_{j^*}}=V_{j^*}$ as is to be expected i.e. there is no point in attacking an item that already has the maximum rating in the target cluster. By lemma \ref{termc}, term $c$ results in a column vector containing identical values in each row. In particular, we have, 

\begin{Lemma}\label{termc}
Given a column vector $y \in \mathbb{R}^{m \times 1}$  with constant values, all rows of column vector $c$ result in an identical positive constant.
\end{Lemma}

\begin{Proof}
 We know that any $i^{th}$ column in $X$ can be written as $x_i=\tilde{U}_t$. Then we can write for any $i^{th}$ row of term $c=y-X^TV_{j^*}$
 \begin{align*}
    y_i-x_i^TV_{j^*} &= y_i-\tilde{U}_t^TV_{j^*} 
\end{align*}
$\tilde{U}_t^TV_{j^*}$ is the predicted rating of item $j^*$ in cluster $t$ before attack and is a constant for all users belonging to cluster $t$. Also $y_i=5$ in our set-up, thus rows of $y-X^TV_{j^*}$ will result in the same non-zero positive value.
\end{Proof}

Now that we have terms $b,c$, we proceed to find the values of the column vector resulting from the product of terms $b$ and $c$.
\begin{Lemma}\label{bc}
 Given $m \times m$ matrix $b=(I+X^TA^{-1}X)^{-1}$ and $m \times 1$ matrix $c=y-X^TV_{j^*}$, the $m \times 1$ column vector resulting from their product has same value for all rows.    
\end{Lemma}
\begin{Proof}
We can write $bc$ as
\begin{align*}
     bc&= b(y-X^TV_{j^*})= by-b(X^TV_{j^*})= by-b\left([\tilde{U}_t,\tilde{U}_t,\cdots,\tilde{U}_t]^TV_{j^*}\right)\\
           &\text{    We have
                $[\tilde{U}_t,\tilde{U}_t,\cdots,\tilde{U}_t]^TV_{j^*}=[\tilde{U}_t^TV_{j^*},\tilde{U}_t^TV_{j^*},\cdots,\tilde{U}_t^TV_{j^*}]^T
                   $ So we can write}\\
     bc&=by-b[\tilde{U}_t^TV_{j^*},\tilde{U}_t^TV_{j^*},\cdots,\tilde{U}_t^TV_{j^*}]^T \\
                &\text{  Using lemma \ref{termb},  $b$ is a matrix with all diagonal values identical and all non-diagonal values identical}\\
                &\text{  and we know $\tilde{U}_t^TV_{j^*}$ is a constant.}\\
     bc&=[K,K,\cdots,K]^T \text{ for positive constant $K$}
 \end{align*}
\end{Proof}

Now we have $\hat{V}_{j^*}-V_{j^*}=A^{-1}X(bc)$ where $bc=[K,K,\cdots,K]^T$ for positive constant $K$. 
From lemma \ref{Xbc}, computing the product $X(bc)$ results in a $\mathbb{R}^{d \times 1}$ vector with values as $m \times K \times \tilde{U}_{t}$.

\begin{Lemma}\label{Xbc}
   Given $X \in \mathbb{R}^{d \times m}$ with $i^{th}$ column vector $x_i$ as $\tilde{U}_t$ and $bc=[K,K,\cdots,K]^T$, we can write $X(bc)=m \times K \times \tilde{U}_t$.
\end{Lemma}
\begin{Proof}
\begin{align*}
   \text{We can write $X(bc)$ as,}\\
    X(bc)&=[x_1,x_2,\cdots,x_m][K,K,\cdots,K]^T\\
         &\text{Using $x_i=\tilde{U}_t$ where $\tilde{U}_t=[u^t_1,u^t_2,\cdots,u^t_d]^T$}\\
         &=\begin{bmatrix}
              u^{t}_{1} & u^{t}_{1} &\cdots, &u^{t}_{1}\\
              u^{t}_{2} & u^{t}_{2} &\cdots, &u^{t}_{2}\\
              \vdots \\
              u^{t}_{d} & u^{t}_{d} &\cdots, &u^{t}_{d} 
            \end{bmatrix}            
            \begin{bmatrix}
                K\\
                K\\
                \vdots\\
                K
            \end{bmatrix}
           =  \begin{bmatrix}                
           u^{t}_{1}K+u^{t}_{1}K+\cdots+u^{t}_{1}K\\  u^{t}_{2}K+u^{t}_{2}K+\cdots+u^{t}_{2}K\\
                \vdots\\           
           u^{t}_{d}K+u^{t}_{d}K+\cdots+u^{t}_{d}K
            \end{bmatrix}\\
            &=                
            [mKu^{t}_{1},  
            mKu^{t}_{2},
                \cdots,       
            mKu^{t}_{d}]^T=m \times K \times \tilde{U}_t
\end{align*}    
\end{Proof}

Equation \ref{Uconst_iter} can now be written as 
$\hat{V_{j^*}}-V_{j^*} = m \times K \times \left(A^{-1}\tilde{U}_t\right)$ 

\subsection{Proof of Lemma \ref{eqn7_behaviour}}
\begin{Definition}\label{identical_columns}
Any matrix $M \in \mathbb{R}^{d \times n}$ with identical columns will satisfy $MM^T=nmm^T$ where $m$ is a column of $M$. Similarly we can write $\sum_{i \in \mathcal{U}(j)}U_iU_i^T=n\tilde{U}_g\tilde{U}_g^T$ if the set of $n$ users in $\mathcal{U}(j)$  belong to same cluster $g$ with weight vector $\tilde{U}_g$\end{Definition}

\begin{restatable}{Lemma}{AUbehaviour}\label{eqn7_behaviour}
 Given a target item $j^*$ such that $A^{-1}=\left(\sum_{i \in \mathcal{U}(j^*)}{U_iU_i^T+\lambda I}\right)^{-1}$ where $\mathcal{U}(j^*)$ is the set of $n$ users who rated item $j^*$. Let a block of fake users with feature vectors in $X \in \mathbb{R}^{d \times m}$ enter to increase the predicted rating of item $j^*$ in a cluster $t$ with $X=[\tilde{U}_t,\tilde{U}_t,\cdots,\tilde{U}_t]$ where $\tilde{U}_t=[u^t_1,u^t_2,\cdots,u^t_d]^T$, then
 \newline 1) If the $n$ users belong to target cluster $t$, all terms  $a_{ki}u^t_i$ of any row $k$ of matrix $A^{-1}\tilde{U}_t$  (equation \ref{feature_update}) will have the same sign and the sign will be opposite to $sgn(a_{kk}u^t_k)$. 
\newline 2) If the $n$ users are not exclusive to the target cluster,  all terms  $a_{ki}u^t_i$ of any row $k$ of the matrix $A^{-1}\tilde{U}_t$ ( equation \ref{feature_update}) need not have the same sign and the sign does not depend exclusively on $sgn(a_{kk}u^t_k)$. 
\end{restatable}

\begin{Proof}

 Let target item be rated by all users belonging to target cluster. By Definition \ref{identical_columns}, we have,
 $$A^{-1}=\left(\sum_{i \in \mathcal{U}(j^*)}{U_iU_i^T+\lambda I}\right)^{-1}=(\lambda I+n\tilde{U}_t\tilde{U}_t^T)^{-1}=\frac{1}{n}\left(\frac{\lambda}{n}I+\tilde{U}_t\tilde{U}_t^T\right)^{-1}$$
 To find $A^{-1}$, we apply Sherman-Morrison formula (see Definition \ref{sherman_morrison}) to the inverse term
 \begin{align*}
   \left(\frac{\lambda}{n}I+\tilde{U}_t\tilde{U}_t^T\right)^{-1}&=\frac{n}{\lambda}I-\frac{\frac{n}{\lambda}I\tilde{U}_t\tilde{U}_t^T\frac{n}{\lambda}I}{1+\tilde{U}_t^T\frac{n}{\lambda}I\tilde{U}_t}=\frac{n}{\lambda}I-\frac{\left(\frac{n}{\lambda}\right)^2\tilde{U}_t\tilde{U}_t^T}{1+\frac{n}{\lambda}\tilde{U}_t^T\tilde{U}_t}=\frac{n}{\lambda}I-\frac{\frac{n}{\lambda}\tilde{U}_t\tilde{U}_t^T}{\frac{\lambda}{n}+\tilde{U}_t^T\tilde{U}_t}\\
               &=\frac{n}{\lambda}I-\frac{\frac{n}{\lambda}{\tilde{U}_t\tilde{U}_t^T}}{C} \text{   where denominator $C=\frac{\lambda}{n}+\tilde{U}_t^T\tilde{U}_t$ is a positive constant}\\
               \end{align*}
    Then we can write $A^{-1}$ as
    \begin{align*}
   A^{-1}&=\frac{1}{n}\left(\frac{\lambda}{n}I+\tilde{U}_t\tilde{U}_t^T\right)^{-1}=\frac{1}{\lambda}\left(I-\frac{\tilde{U}_t\tilde{U}_t^T}{C}\right)   \label{Ainv}
\end{align*}
\textbf{Case 1}: Now that we have matrix $A^{-1}$ for the case when true users come exclusively from the target cluster, we calculate $A^{-1}\tilde{U}_t$,
\begin{align*}
   A^{-1}\tilde{U}_t &=\frac{1}{\lambda}\begin{bmatrix}
                1-\frac{(u^t_1)^2}{C} & -\frac{u^t_1u^t_2}{C} & \cdots &-\frac{u^t_1u^t_d}{C}\\
                -\frac{u^t_2u^t_1}{C}  & 1-\frac{(u^t_2)^2}{C} & \cdots &-\frac{u^t_2u^t_d}{C}\\
                \vdots\\ 
                -\frac{u^t_du^t_1}{C} &-\frac{u^t_du^t_2}{C}  &\cdots & 1-\frac{(u^t_d)^2}{C}
               \end{bmatrix} 
               \begin{bmatrix}
                u^t_1\\
                u^t_2\\
                \vdots\\
                u^t_d
               \end{bmatrix}=
                \frac{1}{\lambda}\begin{bmatrix}
                u^t_1\left(1-\frac{(u^t_1)^2}{C}\right)+\sum_{i=1,i \ne 1}^{d}-u^t_1\frac{(u^t_i)^2}{C}\\
                u^t_2\left(1-\frac{(u^t_2)^2}{C}\right)+\sum_{i=1,i \ne 2}^{d}-u^t_2\frac{(u^t_i)^2}{C}\\
                \vdots\\
                u^t_d\left(1-\frac{(u^t_d)^2}{C}\right)+\sum_{i=1,i \ne d}^{d}-u^t_d\frac{(u^t_i)^2}{C}\\
                \end{bmatrix}
\end{align*}
Recall equation \ref{feature_update} that gives the $k^{th}$ row of $A^{-1}\tilde{U}_t$. 
Here the $k^{th}$ row $a_{kk}u^{t}_{k}+\sum_{i=1,i\ne k}^{d} a_{ki}u^{t}_{i} =u^t_k\left(1-\frac{(u^t_k)^2}{C}\right)+\sum_{i=1,i \ne k}^{d}-u^t_k\frac{(u^t_k)^2}{C}$. Note that the first term $u^t_k\left(1-\frac{(u^t_k)^2}{C}\right)$ has the sign of $u^t_k$ since $\frac{(u^t_k)^2}{C}<1$. Consider the terms in the summation i.e. $a_{ki}u^{t}_{i}= -u^t_k\frac{(u^t_i)^2}{C}$. Here $\frac{(u^t_i)^2}{C}$ is positive because the numerator is a square and the denominator is a positive constant. So each term $-u^t_k\frac{(u^t_i)^2}{C}$ has sign $-sgn(u^t_k)$ or in other words opposite to the sign of the first term. Therefore, for case $1$, all terms  $a_{ki}u^t_i$ of any row $k$ of the matrix $A^{-1}\tilde{U}_t$  (equation \ref{feature_update})  has the same sign and the sign is opposite to $sgn(a_{kk}u^t_k)$.  

\vspace{2mm}
\textbf{Case 2}: Following similar steps,  $A^{-1}\tilde{U}_t$ when true ratings come exclusively from any cluster $g \ne t$ and for $C=\frac{\lambda}{n}+\tilde{U}_g^T\tilde{U}_g$ is given by.
\begin{align*}
   \hat{A}^{-1}\tilde{U}_t &=\frac{1}{\lambda}\begin{bmatrix}
                1-\frac{(u^g_1)^2}{C} & -\frac{u^g_1u^g_2}{C} & \cdots &-\frac{u^g_1u^g_d}{C}\\
                -\frac{u^g_2u^g_1}{C}  & 1-\frac{(u^g_2)^2}{C} & \cdots &-\frac{u^g_2u^g_d}{C}\\
                \vdots\\ 
                -\frac{u^g_du^g_1}{C} &-\frac{u^g_du^g_2}{C}  &\cdots & 1-\frac{(u^g_d)^2}{C}
               \end{bmatrix} 
               \begin{bmatrix}
                u^t_1\\
                u^t_2\\
                \vdots\\
                u^t_d
               \end{bmatrix}=              
                \frac{1}{\lambda}\begin{bmatrix}
                u^t_1\left(1-\frac{(u^g_1)^2}{C}\right)+\sum_{i=1,i \ne 1}^{d}-u^t_i\frac{u^g_1u^g_i}{C}\\
                u^t_2\left(1-\frac{(u^g_2)^2}{C}\right)+\sum_{i=1,i \ne 2}^{d}-u^t_i\frac{u^g_2u^g_i}{C}\\
                \vdots\\
                u^t_d\left(1-\frac{(u^g_d)^2}{C}\right)+\sum_{i=1,i \ne d}^{d}-u^t_i\frac{u^g_2u^g_i}{C}\\
                \end{bmatrix}
\end{align*}
Similar to previous case, first term in $k^{th}$ row of $A^{-1}\tilde{U}_t$ has the sign of $u^t_k$. The terms in summation $a_{ki}u^g_i=-u^t_i\frac{u^g_ku^g_i}{C}$, has its sign depend on $sgn(u^t_iu^g_ku^g_i)$. So unlike case 1, term $a_{ki}u^g_i$ may have any sign.

\subsubsection*{\textbf{When true ratings come from a combination of clusters:}} Consider a simple extension where $n_1$ users of target cluster $t$ and $n_2$ users of cluster $g$ provide ratings. Given $n_1+n_2=n$, we have,
\begin{align*}
 A^{-1}&=(\lambda I+n_1\tilde{U}_t\tilde{U}_t^T+n_2\tilde{U}_g\tilde{U}_g^T)^{-1}=(M+n_2\tilde{U}_g\tilde{U}_g^T)^{-1}\\ 
&=M^{-1}-\frac{M^{-1}\tilde{U}_g\tilde{U}_g^TM^{-1}}{\frac{1}{n_2}+\tilde{U}_g^TM^{-1}\tilde{U}_g} \text{ where $M^{-1}=\left(\lambda I+n_1\tilde{U}_t\tilde{U}_t^T\right)^{-1}$ (given in case $1$)}    
\end{align*}
Caclulating $A^{-1}\tilde{U}_t$
\begin{align*}
    A^{-1}\tilde{U}_t
          &=M^{-1}\tilde{U}_t-\frac{M^{-1}\tilde{U}_g\tilde{U}_g^TM^{-1}\tilde{U}_t}{\frac{1}{n_2}+\tilde{U}_g^TM^{-1}\tilde{U}_g}=M^{-1}(\tilde{U}_t-\beta\tilde{U}_g) \text{ where constant $\beta=\frac{ \tilde{U}_g^TM^{-1}\tilde{U}_t}{\frac{1}{n_2}+\tilde{U}_g^TM^{-1}\tilde{U}_g}$}\\
          &= \frac{1}{\lambda}\begin{bmatrix}
                (u^t_1-\beta u^g_1)\left(1-\frac{(u^t_1)^2}{C}\right)+\sum_{i=1,i \ne 1}^{d}-\frac{-u^t_1u^t_i}{C}(u^t_i-\beta u^g_i)\\
                (u^t_2-\beta u^g_2)\left(1-\frac{(u^t_2)^2}{C}\right)+\sum_{i=1,i \ne 2}^{d}-\frac{-u^t_2u^t_i}{C}(u^t_i-\beta u^g_i)\\
                \vdots\\
                (u^t_d-\beta u^g_d)\left(1-\frac{(u^t_d)^2}{C}\right)+\sum_{i=1,i \ne d}^{d}-\frac{-u^t_du^t_i}{C}(u^t_i-\beta u^g_i)\\
                \end{bmatrix}
\end{align*}
For any row $k$ in the resultant matrix, while the first term will have $sgn(u^t_k-\beta u^g_k)$,  each term in the summation given by $\frac{-u^t_ku^t_i}{C}(u^t_i-\beta u^g_i)$ has its sign depending on many factors. Similarly, we can extend the analysis to true ratings coming from any combination of clusters by finding the inverse recursively per cluster rating block using the Sherman-Morrison update. 

Thus except for case 1,  for any combination of cluster-wise distribution of true ratings, all terms  $a_{ki}u^t_i$ of any row $k$ of the matrix $A^{-1}\tilde{U}_t$ ( equation \ref{feature_update}) need not have the same sign and the sign does not depend exclusively on $sgn(a_{kk}u^t_k)$.
\end{Proof}

\subsection{Proof of Theorem \ref{thm:targetratingslemma}}
\begin{Proof}
Given $A^{-1}$, let $N$ users from target cluster provide ratings.
Then for $\epsilon>0$ and $N\ge\frac{1}{\epsilon}$, we update the inverse by Sherman-Morrison formula (Definition \ref{sherman_morrison})
 \begin{align*}
     \hat{A}^{-1} &=(A+N\tilde{U}_t\tilde{U}_t^T)^{-1} = A^{-1}-\frac{A^{-1}\tilde{U}_t\tilde{U}_t^TA^{-1}}{\frac{1}{N}+\tilde{U}_t^TA^{-1}\tilde{U}_t}
     =A^{-1}-\frac{NA^{-1}\tilde{U}_t\tilde{U}_t^TA^{-1}}{1+N\tilde{U}_t^TA^{-1}\tilde{U}_t} 
 \end{align*}

Now that we have $\hat{A}^{-1}$, we calculate $\hat{A}^{-1}\tilde{U}_t$ of equation \ref{A_inverseU}.
  \begin{align*}
     \hat{A}^{-1}\tilde{U}_t = \left(A^{-1}-\frac{NA^{-1}\tilde{U}_t\tilde{U}_t^TA^{-1}}{1+N\tilde{U}_t^TA^{-1}\tilde{U}_t}\right)\tilde{U}_t =\left(\frac{A^{-1}\tilde{U}_t}{1+N\tilde{U}_t^TA^{-1}\tilde{U}_t}\right)
 \end{align*}
 We see that as $\epsilon \to 0$, $N \to \infty$,  causing $\hat{A}^{-1}\tilde{U}_t$ to converge to zero vector.   

Now suppose, instead of the target cluster, users from any non-target cluster $g$ provide the $N$ ratings; then we have
   \begin{align*}
     \hat{A}^{-1}\tilde{U}_t &\le \left(A^{-1}-\frac{A^{-1}\tilde{U}_g\tilde{U}_g^TA^{-1}}{\epsilon+\tilde{U}_g^TA^{-1}\tilde{U}_g}\right)\tilde{U}_t=\left(A^{-1}\tilde{U}_t-\frac{A^{-1}\tilde{U}_g(\tilde{U}_g^TA^{-1}\tilde{U}_t)}{\epsilon+\tilde{U}_g^TA^{-1}\tilde{U}_g}\right)=\left(\epsilon A^{-1}\tilde{U}_t +C_1A^{-1}\tilde{U}_t - C_2A^{-1}\tilde{U}_g \right)
 \end{align*}
 where constants $C_1=\frac{\tilde{U}_g^TA^{-1}\tilde{U}_g}{\epsilon+\tilde{U}_g^TA^{-1}\tilde{U}_g}$ and  $C_2=\frac{\tilde{U}_g^TA^{-1}\tilde{U}_t}{\epsilon+\tilde{U}_g^TA^{-1}\tilde{U}_g}$.
 As $C_1,C_2$ are non-zero constants with $sgn(C_1)$ as positive and $sgn(C_2)$ depending on $sgn(\tilde{U}_g^TA^{-1}\tilde{U}_t)$, $\hat{A}^{-1}\tilde{U}_t$ might not converge to a zero vector.
\end{Proof}

\section{Additional Experimental Results}
 We look at a slightly modified hit-rate definition, defining an item to belong to the recommended list if it has a predicted rating $>=4$. If the target item after the attack satisfies this condition, we consider it a 'Hit.' i.e, it may be more likely for the target item to be now recommended to users of the target cluster. We measure the Average Hits after the attack for the target item and show the results when the adversary targets cluster 2. Results when targeting other clusters are similar and so are not reported separately.
\subsection{Fix $V$, Update $U$}
From Table \ref{tab:HR_U_vary},  Average-Hits over all the target items calculated is $0$ for both datasets for increasing ratios of $\frac{n}{m}$, indicating that the attack is not able to push the rating high enough to be included in the recommended list of items.  i.e, the results of the attack here, do not increase the predicted rating to $\ge 4$. This supports our results in the main paper (figure \ref{fig:Vconst_vary_Nt}), where the maximum mean relative change in rating observed for $\frac{n}{m}=0.5$ was $< 0.30$ i.e. for both datasets, predicted rating increases only up to $30\%$ of maximum deviation possible. 
\begin{table}
\subfloat[Movielens]{
\begin{tabular}{|l|l|l|l|l|l|l|}
\hline
 $\frac{n}{m}$/Avg Hits   & Before & After       \\ \hline
$0.5$ & 0 & 0  \\ \hline
$1.2$    & 0 & 0 \\ \hline
$2.0$  & 0 & 0  \\ \hline
\end{tabular}
}
\qquad
\subfloat[Goodreads]{
\begin{tabular}{|l|l|l|l|l|l|l|}
\hline
 $\frac{n}{m}$/Avg Hits   & Before & After       \\ \hline
$0.5$ & 0 & 0  \\ \hline
$1.2$    & 0 & 0 \\ \hline
$2.0$  & 0 & 0  \\ \hline
\end{tabular}
}
\caption{Average Hits for Movielens dataset and Goodreads dataset reported when target cluster is $t=2$ for the cluster-wise distribution of true users $250-250-n-250$ }\label{tab:HR_U_vary}
\end{table}

\subsection{Fix $U$, Update $V$}
\subsubsection*{Varying Ratio of Target Item True Ratings ($N_t$) to Fake Ratings ($N_f$) in Target Cluster}
Table \ref{tab:HR_dist1} reports the average hits for the target item for increasing $N_t$ from the target cluster given $N_f$ fake ratings. We can see how the average hit is highest for $\frac{N_t}{N_f}=0.05,0.5,1$ and drops to zero afterward. This is in agreement with the observations from Figure \ref{fig:target item ratings}, where $\frac{N_t}{N_f}=0.05,0.5,1$  showed a significant mean shift in the predicted rating $>=0.5$  after the attack compared to the mean shift in the predicted rating $<0.4$ for $\frac{N_t}{N_f}>1$. Therefore, the attack increases the predicted rating to $\ge 4$, raising the chances of being recommended to users of the target cluster, for decreasing $\frac{N_t}{N_f}$. 
\begin{table}
\subfloat[Movielens]{
\begin{tabular}{|l|l|l|l|l|l|l|}
\hline
 $\frac{N_t}{N_f}$/Avg Hits   & Before & After       \\ \hline
$0.05$ & 0 & 1  \\ \hline
$0.5$    & 0 & 1 \\ \hline
$1$    & 0 & 1 \\ \hline
$1.75$    & 0 & 0 \\ \hline
$2.5$  & 0 & 0  \\ \hline
\end{tabular}
}
\qquad
\subfloat[Goodreads]{
\begin{tabular}{|l|l|l|l|l|l|l|}
\hline
 $\frac{N_t}{N_f}$/Avg Hits   & Before & After       \\ \hline
$0.05$ & 0 & 1  \\ \hline
$0.5$    & 0 & 1 \\ \hline
$1$    & 0 & 1 \\ \hline
$1.75$    & 0 & 0 \\ \hline
$2.5$  & 0 & 0  \\ \hline
\end{tabular}

}
\caption{Average Hits for Movielens dataset and Goodreads dataset reported when target cluster is $t=2$ for the rating distribution of type $0-0-N_t-0$.}\label{tab:HR_dist1}
\end{table}

\subsubsection*{ Varying Ratio of Target Item True Ratings ($N_t$) to Fake Ratings ($N_f$) in Non-Target Clusters}
Table \ref{tab:HR_dist2} reports the average hits for target item for each dataset. We can see how the average hit remains high for increasing $\frac{N_t}{N_f}$. This agrees with the observations from Figure \ref{fig:r5_in_ng}, where increasing $\frac{N_t}{N_f}$ showed larger mean shift in predicted rating of $>0.95$  after the attack increasing the predicted rating to $\ge 4$, thus raising the chances of being recommended to users of target cluster. 

\begin{table}[H]
\subfloat[Movielens]{
\begin{tabular}{|l|l|l|l|l|l|l|}
\hline
 $\frac{N_t}{N_f}$/Avg Hits   & Before & After       \\ \hline
$0.05$ & 0 & 1  \\ \hline
$0.5$  & 0 & 1  \\  \hline
$1$    & 0 & 1 \\ \hline
$1.75$  & 0 & 1  \\  \hline
$2.5$  & 0 & 1  \\ \hline
\end{tabular}
}
\qquad
\subfloat[Goodreads]{
\begin{tabular}{|l|l|l|l|l|l|l|}
\hline
 $\frac{N_t}{N_f}$/Avg Hits   & Before & After       \\ \hline
$0.05$ & 0 & 1  \\ \hline
$0.5$  & 0 & 1  \\  \hline
$1$    & 0 & 1 \\ \hline
$1.75$  & 0 & 1  \\  \hline
$2.5$  & 0 & 1  \\ \hline
\end{tabular}
}
\caption{Average Hits for Movielens dataset and Goodreads dataset reported when target cluster is $t=2$ for the rating distribution of type $N_t-N_t-0-N_t$.}\label{tab:HR_dist2}
\end{table}

\subsection{Results on ML-1M Dataset}\label{ML-1M dataset}

\begin{figure}[H]
\centering
\subfloat[]{
\includegraphics[width=0.35\columnwidth]{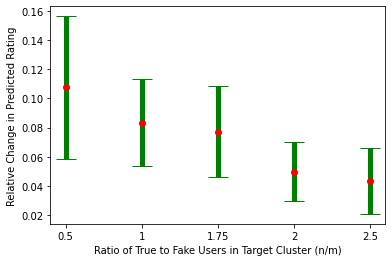}
}
\subfloat[]{
\includegraphics[width=0.35\columnwidth]{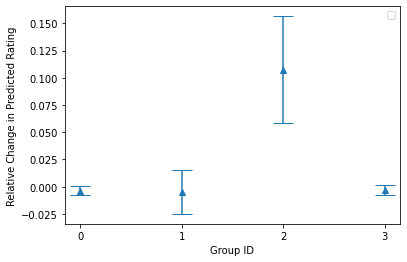}
}\\

\caption{Plot comparing the relative change in the predicted rating a) against increasing ratio $\frac{n}{m}$ in the target cluster 2 b) against other user clusters for $\frac{n}{m}=0.5$ when targeting cluster 2
}\label{fig:ML_med_U}
\end{figure}

\begin{figure}[H]
\centering
\subfloat[]{
\includegraphics[width=0.35\columnwidth]{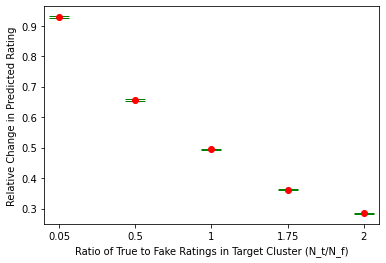}
}
\subfloat[]{
\includegraphics[width=0.35\columnwidth]{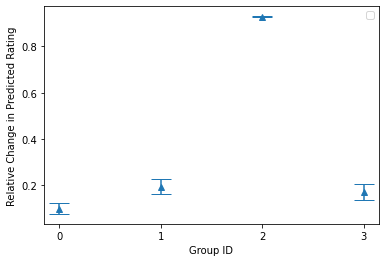}
}\\

\caption{Plot comparing the relative change in predicted rating a) against increasing ratio $\frac{N_t}{N_f}$ in the target cluster 2 b) against other user clusters for $\frac{N_t}{N_f}=0.05$
}\label{fig:ML_med_v}
\end{figure}

Figures \ref{fig:ML_med_U} and \ref{fig:ML_med_v} illustrates results in ML-1M dataset. Figure \ref{fig:ML_med_U} demonstrates the effect of attack when only $U$ is updated. As observed in the other datasets, Figure \ref{fig:ML_med_U}(b) illustrates that the changes in the target group centre $\tilde{U}_t$ is negligible causing less than $15\%$ change in predicted rating in the target cluster and Figure \ref{fig:ML_med_U}(a) shows that increase in the number of true users relative to fake users diminishes the effect of attack the in target cluster gradually.

Figure \ref{fig:ML_med_v} demonstrates the effect of attack when only $V$ is updated.  
We can see from Figure \ref{fig:ML_med_v}(a) that the relative change in rating after the attack, due to a shift in item vector $V$, reduces from $\sim 0.9$ to $\sim 0.3$ with increasing number of true ratings in the targeted cluster. Figure \ref{fig:ML_med_v}(b) demonstrates that the attack leaks to non-targeted clusters as well, similar to our observations in the other datasets.

\subsection{Demoting Target Item Rating}
Recall the definition of the evaluation metric 'Relative Change in Rating' in equation \ref{eq:metric}. For visualising the demotion of items,  instead of 5, we assign 1 as the minimum rating that can be given to the target item in the datasets that we consider. Thus equation \ref{eq:metric} can be updated as

\begin{align} 
\text{Relative Change in Rating}_g = \frac{\mu_f(g,i) - \mu_o(g,i)}{|1-\mu_o(g,i)|}
\end{align}
where $\mu_f(u, i)$ is the predicted rating of target item $i$ of a user in cluster $g$ after the attack, $\mu_o(g, i)$ is the predicted rating of the target item $i$ of a user in cluster $g$ before the attack.   

\begin{figure}[H]
\centering
\subfloat[]{
\includegraphics[width=0.35\columnwidth]{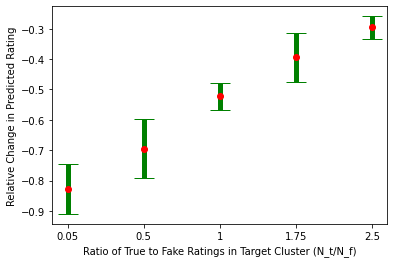}
}

\caption{Plot comparing the relative change in predicted rating against increasing ratio $\frac{N_t}{N_f}$ in the target cluster in ML-1M dataset
}\label{fig:V_demote}
\end{figure}

Consider Figure \ref{fig:V_demote}, which shows the relative change in rating for increasing ratio of $\frac{N_t}{N_f}$ in the target cluster. The negative change in rating implies that the rating of an item reduces from its original value following the attack.

When $\frac{N_t}{N_f}$ equals 0.05, we observe that the rating of the target item decreases to approximately 85\% of the maximum possible deviation (by rating value reduction) for the item. Similar to what we observed in the case of promotion, this effect diminishes as the ratio increases. Consequently, these observations and results can be extended to scenarios involving item demotion among specific user clusters.

\subsection{Results for Standard MF Methods}\label{standardMF}

\begin{figure}[H]
\centering
\subfloat[]{
\includegraphics[width=0.35\columnwidth]{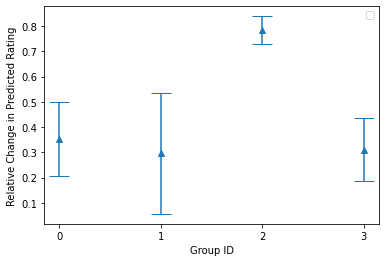}
}
\subfloat[]{
\includegraphics[width=0.35\columnwidth]{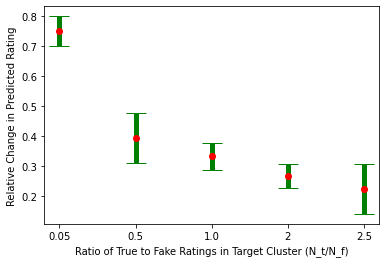}
}\\

\caption{Plot comparing the relative change in predicted rating  a) against other user clusters for $\frac{N_t}{N_f}=0.05$  b) against increasing ratio $\frac{N_t}{N_f}$ in the target cluster
under standard MF update}\label{fig:std MF results}
\end{figure}

Figures \ref{fig:std MF results} demonstrate the key results of our study under the standard MF update in the ML-100k dataset. \footnote{The results follow a similar trend in other datasets and so are not reported separately} Let us recall, that it is the changes to item vector $V_{j^*}$ of the target item $j^*$ that cause the most changes to the item's predicted rating and is also responsible for propagating the attack across all user clusters. These changes to $V_{j^*}$ are guaranteed to be reduced if we increase ratings to the target item in the target cluster.

Figure  \ref{fig:std MF results}(a) shows  that the attack effects leak to non-target clusters as demonstrated in Section \ref{result_varyV} due to the changes in item vector $V_{j^*}$ of the target item $j^*$. In Figure \ref{fig:std MF results}(b), we demonstrate the change in target item predicted rating as the ratio $\frac{N_t}{N_f}$ increases. As observed in our previous findings, the impact of attacks due to shifts to $V_{j^*}$  diminishes as the number of true ratings in the target cluster increases.

\end{document}